\documentclass[traditabstract]{aa} 
\usepackage{epsfig}
\usepackage{amsmath}
\usepackage{color}
\usepackage{txfonts}
\usepackage[mediumspace,mediumqspace,Grey,squaren]{SIunits}
\usepackage{multirow}

\newcommand{\HI}{H\,{\sc {i}}~}

\newcommand{\Msold}{M$_{\odot}$\,yr$^{-1}$}
\newcommand{\Mdot}{10$^{-7}$\,M$_{\odot}$\,yr$^{-1}$}

\newcommand{\lsim}{~\rlap{$<$}{\lower 1.0ex\hbox{$\sim$}}}
\newcommand{\gsim}{~\rlap{$>$}{\lower 1.0ex\hbox{$\sim$}}}

\begin{document}
 \title{$^{12}$CO emission from EP Aqr: Another example of an axi-symmetric 
AGB wind?\thanks{Based on observations carried out 
with the IRAM Plateau-de-Bure Interferometer and the IRAM 30-m telescope. 
IRAM is supported by INSU/CNRS (France), MPG (Germany) and IGN (Spain).}}

   \author{P. T. Nhung\inst{1,2},
           D. T. Hoai\inst{1,2},  
	   J. M. Winters\inst{3},
	   T. Le\,Bertre\inst{1},
           P. N. Diep\inst{2},\\
           N. T. Phuong\inst{2},
           N. T. Thao\inst{2},
           P. Tuan-Anh\inst{2}
          \and P. Darriulat\inst{2}
          }

   \institute{LERMA, UMR 8112, CNRS \& Observatoire de Paris/PSL, 
	      61 av. de l'Observatoire, F-75014 Paris, France 
         \and
              Department of Astrophysics, Vietnam National Satellite Center, VAST,
              18 Hoang Quoc Viet, Ha Noi, Vietnam
         \and
              IRAM, 300 rue de la Piscine, Domaine Universitaire, 
	      F-38406 St. Martin d'H\`eres, France
             }

   \date{\today}

   \titlerunning{$^{12}$CO emission from EP Aqr}
   \authorrunning{P. T. Nhung, D. T. Hoai, J. M. Winters, et al.}


  \abstract 
{The CO(1-0) and (2-1) emission of the circumstellar envelope of the
Asymptotic Giant Branch (AGB) star EP Aqr has been observed in
2003 using the IRAM Plateau-de-Bure Interferometer and in 2004 using
the IRAM 30-m telescope at Pico Veleta. The line profiles reveal the
presence of two distinct components centered on the star velocity, a
broad component extending up to ${\sim}10$ km\,s$^{-1}$ and a narrow
component indicating an expansion velocity of only ${\sim}2$
km\,s$^{-1}$. An early analysis of these data was performed under the
assumption of isotropic winds. The present study revisits this
interpretation by assuming instead a bipolar outflow nearly aligned
with the line of sight. A satisfactory description of the observed
flux densities is obtained with a radial expansion velocity increasing
from ${\sim}2$ km\,s$^{-1}$ at the equator to ${\sim}10$ km\,s$^{-1}$
near the poles. The mass loss rate is ${\sim}1.8\,10^{-7}$
\Msold. The angular aperture of the bipolar outflow is
${\sim}45^\circ$ with respect to the star axis, which makes an angle
of ${\sim}13^\circ$ with the line of sight. A detailed study of the CO(1-0) to CO(2-1)
flux ratio reveals a significant dependence of the temperature on the
star latitude, smaller and steeper at the poles than at the equator at
large distances from the star ($>2''\equiv1.0\times10^{-3}$\,pc). 
Under the hypothesis of radial
expansion of the gas and of rotation invariance about the star axis,
the effective density has been evaluated in space as a function of
star coordinates (longitude, latitude and distance from the star).
Evidence is found for an enhancement of the effective density 
in the northern hemisphere of the star at
angular distances in excess of ${\sim}3''$ and covering the whole
longitudinal range. The peak velocity of the narrow component is
observed to vary slightly with position on the sky, a variation
consistent with the model and understood as the effect of the
inclination of the star axis with respect to the line of sight. 
This variation is inconsistent with the assumption of a spherical wind
and strengthens our interpretation in terms of an axisymmetric outflow.
While the phenomenological model presented here reproduces well 
the general features of the observations, not only qualitatively but also
quantitatively, significant  differences are also revealed, which would
require a better spatial resolution to be properly described and
understood.}


   \maketitle
%

\section{Introduction}\label{introsec}

EP Aqr is one of the nearest mass losing Asymptotic Giant Branch 
(AGB) stars ({\it d}=114 pc,
van Leeuwen 2007), and as such one of the best characterized objects
of its class. The absence of Technetium in the spectrum (Lebzelter 
\& Hron 1999), and the low value of the $^{12}$C/$^{13}$C abundance ratio 
($\sim$ 10, Cami et al. 2000) indicate that it
is still at the beginning of its evolution on the AGB. Yet Herschel
has imaged a large scale ($2'\times4'$) circumstellar shell at
$\unit{70}\micro\meter$ (Cox et al. 2012) testifying for a relatively
long duration of the mass loss (${>}10^4$ years) and for interaction
with the surrounding interstellar medium. The elongation of the far-infrared 
image is in a direction opposite to the EP Aqr space motion. 
From \HI observations at 21 cm, Le~Bertre \& G\'erard (2004) estimate 
a duration of $\sim$ 1.6$\times$10$^5$ years for the present episode of 
mass loss.

Combined observations have been obtained with the IRAM 30-m telescope
and the Plateau-de-Bure Interferometer in $^{12}$CO(1-0) and
$^{12}$CO(2-1) in 2003 and 2004, improving over earlier observations
(Knapp et al. 1998, Nakashima 2006). The spectra, which reveal a
wind composed of two main components with expansion velocities
${\sim}2$ and ${\sim}10$ km\,s$^{-1}$, were analyzed in terms of
multiple isotropic winds (Winters et al. 2003, 2007). The
interferometer data show an extended source of about 15$''$ (FWHM),
and evidence for a ring structure in CO(2-1). 
However, when assuming multiple isotropic winds, Winters et al. (2007)
did not manage to fully explain the line profiles obtained
at different positions of the spectral maps.

Composite CO line profiles have been observed for several AGB 
sources,  including RS Cnc and X Her (e.g. Knapp et al. 1998). Kahane
\& Jura (1996) interpreted the narrow line component of X Her in terms
of a spherically expanding wind and the broad line component in terms
of a bipolar flow, probably observed at a viewing angle of about
15$^{\circ}$. For RS Cnc, high spatial resolution observations can be
interpreted assuming the same geometry but with a different
inclination of about 40$^{\circ}$ with respect to the line of sight
(Hoai et al. 2014).  These results raise the question whether the
bipolar outflow hypothesis might also apply to EP Aqr.

The present study is therefore revisiting the analysis of
Winters et al. (2007), assuming instead a bipolar outflow
approximately directed along the line of sight.  We exploit the
experience gained in similar studies of other stars, RS Cnc (Hoai et
al. 2014, Nhung et al. 2015) and the Red Rectangle, a post-AGB source, 
which could be interpreted by applying a similar model (Tuan Anh et al. 2015).

The paper is organised as follows: in Sect.~\ref{obssec} we
present reprocessed data on EP Aqr that will be used in this work.
Sect.~\ref{geometrysec} describes in an elementary way the main
features implied by the assumed orientation of EP Aqr, with the
star axis nearly parallel to the line of sight, mimicking spherical
symmetry.  Sect.~\ref{compsec} presents an analysis of the data using
a simple bipolar outflow model that had been developed earlier for
RS Cnc (Hoai et al. 2014, Nhung et al. 2015); the
results of the best fit to the observations is used as a reference in
the following sections. Sect.~\ref{fluxsec} presents a study of the
CO(1-0) to CO(2-1) emission ratio, allowing for an evaluation of the
temperature distribution nearly independent of the model adopted for
the description of the gas density and
velocity. Sect.~\ref{meridplanesec} uses the distribution of the gas
velocity obtained in Sect.~\ref{compsec} to reconstruct directly in
space the effective gas density, without making direct use of the
parametrization of the flux of matter obtained in
Sect.~\ref{compsec}, and providing therefore a consistency check of
the results of the model. Finally, Sect.~\ref{narrowsec} presents a
detailed study of the spatial  variation of the radial velocity of the
narrow line component.  It provides a sensitive check of the bipolar
outflow hypothesis and of its small inclination angle with respect to
the line of sight.

\section{Observations}\label{obssec}
The observations analysed here combine interferometer data obtained
using the Plateau-de-Bure Interferometer with short-spacing data
obtained using the Pico Veleta 30-m telescope. A detailed description
of the original observations is given in Winters et al. (2007).
$^{12}$CO(2-1) and $^{12}$CO(1-0) spectral data have been obtained with
a spatial resolution of ${\sim}1''$ and ${\sim}2''$ respectively and a
spectral resolution of 0.1 km\,s$^{-1}$. At such spatial resolution,
the two spectral components present in the single-dish line spectra are
seen as originating from the same region. The images are virtually
circularly symmetric and display smooth variations with velocity and
projected distance from the star. In Winters et al. (2007) the
observed spectra were interpreted by a succession of spherically
symmetric and short-spaced mass-loss events. These authors noted the
presence of inhomogeneities in the spectral maps indicating a possible
clumpiness of the circumstellar envelope.

\begin{figure}
\centering
\includegraphics[width=8.5cm]{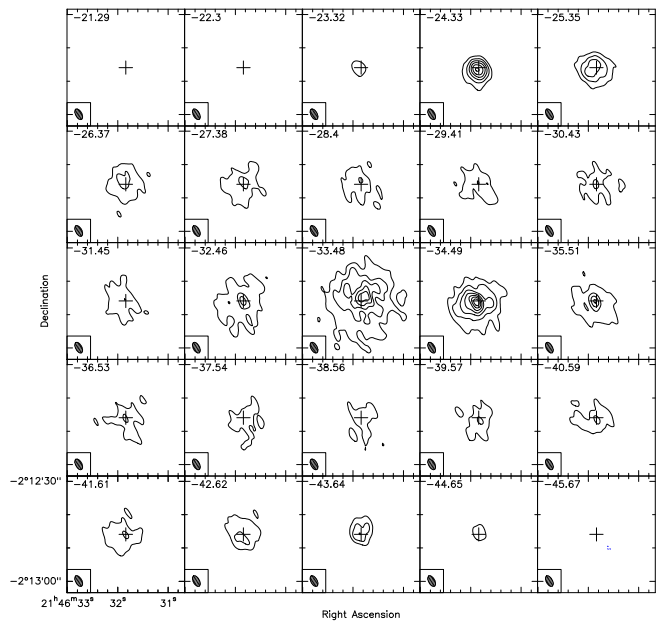}
\caption{Channel maps in the $^{12}$CO(J=1-0) line (smoothed to a 
width of 1\,km\,s$^{-1}$). Contours are plotted at 5, 10, 20, 30, 40\,$\sigma$ 
(1\,$\sigma = 14\,$mJy/beam). 
The synthesized beam is indicated in the lower left.}
  \label{co10mapfig}
\end{figure}

\begin{figure}
\centering
\includegraphics[width=8.5cm]{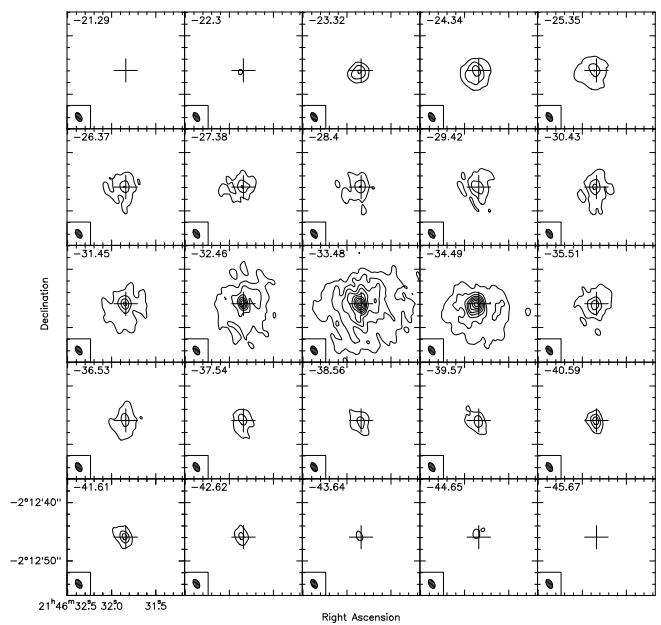}
\caption{Channel maps in the $^{12}$CO(J=2-1) line (smoothed to a
width of 1\,km\,s$^{-1}$). First contour is plotted at 10\,$\sigma$,
the following contours start at 20\,$\sigma$ and are plotted in
20\,$\sigma$ steps (1\,$\sigma = 16\,$mJy/beam). The
synthesized beam is indicated in the lower left.}
  \label{co21mapfig}
\end{figure}

We have reprocessed the original data, correcting for an artifact introduced by
a velocity shift of 0.52\,km\,s$^{-1}$ in the CO(1-0) data and
recentering the maps on the position of the star at epoch 2004.0,  
correcting for its
proper motion as determined by van~Leeuwen (2007). The reprocessed
channel maps are shown in Figs.~\ref{co10mapfig} and
\ref{co21mapfig}.

\section{Observing a star along its symmetry axis}\label{geometrysec}

\begin{figure}
\centering
\includegraphics[height=7.cm,width=7.4cm,trim=1.cm 1.cm 1.cm 1.cm,clip]{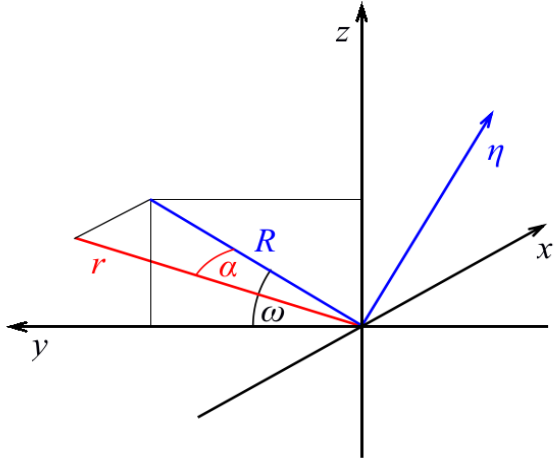}
\caption{Definition of coordinates. The $(y,z)$ coordinates are obtained by 
rotation of angle $\omega$ in the sky plane about the $x$ axis (line of sight 
and star axis, cf. Sect.~\ref{geometrysec}). In the sky plane, 
y is towards East, and z towards North.}
  \label{fig1}
\end{figure}

The high degree of azimuthal symmetry observed in the channel maps of
both CO(1-0) and CO(2-1) emission (see Figs.~\ref{co10mapfig} and
\ref{co21mapfig}, Winters et al.~(2007)) implies that
a bipolar outflow, if present, should have its axis nearly parallel to
the line of sight. In such a configuration, some simple relations
apply between the space coordinates and their
projection on the sky plane. We review them briefly in order to ease
further discussion. We use coordinates (Fig.~\ref{fig1}) centered on
the star with $x$ along the star axis, parallel to the line of sight,
$y$ pointing east and $z$ pointing north.  We assume rotation symmetry
about the $x$ axis and symmetry about the equatorial $(y,z)$ plane of
the star. Defining $\rho$ the gas density, $T$ the gas temperature and
$\boldsymbol{V}$ the gas velocity in the star rest frame, 
having components $V_x$, $V_y$,
$V_z$, the following relations apply at point $(x, y, z)$: $\rho$ and
$T$ are functions of $r=(x^2+y^2+z^2)^\frac{1}{2}$ and of $|\alpha|$
where $\alpha=arctan(x/R)$ is the latitude in the star reference 
system, and $R=(y^2+z^2)^\frac{1}{2}$ is the
projection of $r$ on the sky plane. \\
The velocity $\boldsymbol{V}$
takes a simple form when expressed in terms of meridian coordinates,
$R=y\cos \omega+z\sin \omega$, $x$ and
$\eta=-y\sin\omega+z\cos\omega$, $\omega$ being the star longitude;
its components are then functions of $r$ and $|\alpha|$. In case of
pure radial expansion, $V_\eta=0$, $V_x=xV/r$ and $V_R=RV/r$ while in
case of pure rotation $V_R=V_x=0$.   

In the case of pure radial expansion with a known radial dependence 
of the velocity, the flux density $f$ measured in
a pixel $(y,z)$ at velocity $V_x$ is associated with a well defined
position in space, obtained from
\begin{equation}
x=rV_x/V=(x^2+R^2)^\frac{1}{2}V_x/V = RV_x(V^2-V_x^2)^{-\frac{1}{2}}.
\label{eq1}
\end{equation}
Then, if one knows $V$, one can calculate $V_x$ and therefore its
derivative $dV_x/dx$ at any point in space. Defining an effective
density $\rho_\mathrm{eff}$ as  $\rho_\mathrm{eff}=fdV_x/dx$, such
that the observed flux in a given pixel be
$\int{fdV_x}=\int{\rho_\mathrm{eff}dx}$, one is then able to calculate
it at any point in space. The effective density, defined by this
relation, is the product of the actual gas density and a factor
accounting for the population of the emitting state, the emission
probability and  correcting for the effect of absorption.

\begin{figure}
\centering
\includegraphics[width=8.5cm,trim=0.cm 1.5cm 0.cm 1.cm,clip]{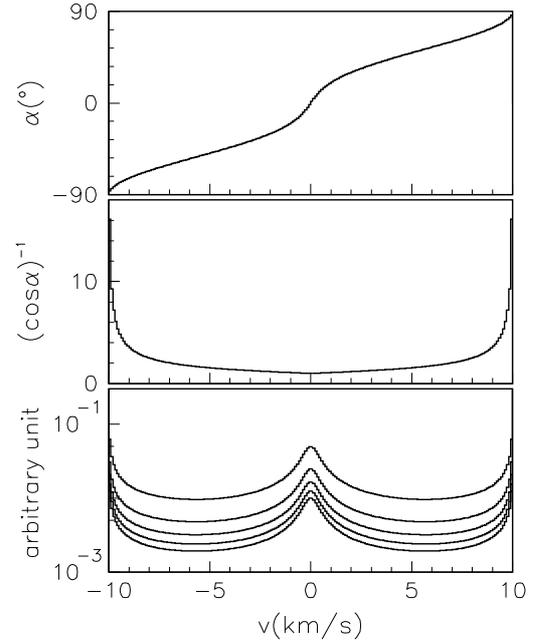}
\caption{Relation between the Doppler velocity $V_x$ and the
star latitude $\alpha$ (cf. Eq.~(\ref{veloeq}), upper panel), the 
ratio $r/R=1/cos\alpha$
(middle panel) for the simple star model described in the text. Lower
panel: velocity spectra obtained for the same model at $R$=1, 2, 3, 4
and 5 (running downward). }
\label{fig2}
\end{figure}

We illustrate the above properties with a simple example, close to the
reality of EP Aqr. We assume exact rotation invariance about the star
axis (parallel to the line of sight), a purely radial wind with a 
velocity that only depends on $\alpha$
\begin{equation}
V=V_0+V_1\sin^2\!\alpha
\label{veloeq}
\end{equation} 
($V_0$ at the equator and $V_0+V_1$ at the poles) 
and an effective density (defined such that its integral over the line
of sight measures the observed flux density) inversely proportional to
$r^2$. From $V_x=V\sin\alpha$ and Eq.~(\ref{veloeq}) 
we obtain $\sin^3\!\alpha+(V_0/V_1)\sin\alpha-(V_x/V_1)=0$. 
This reduced cubic equation is explicitly solvable with
$\sin\alpha=-(q-s)^\frac{1}{3}-(q+s)^\frac{1}{3}$,
$s=(q^2+p^3)^\frac{1}{2}$, $p=\frac{1}{3}V_0/V_1$ and
$q=-\frac{1}{2}V_x/V_1$.  For a given value of the Doppler velocity
$V_x$ one can therefore calculate $\alpha$ independent of $R$: in
any pixel, $V_x$ and $\alpha$ are related in the same way, which is
illustrated in Fig.~\ref{fig2} (upper panel). Similarly, $V_x$ and
$r/R=1/\cos\alpha$ are also related in a universal way, independent
of $R$ (Fig.~\ref{fig2}, middle panel). The resulting velocity spectra
are displayed in Fig.~\ref{fig2} (lower panel) for $R$=1, 2, 3, 4 and
5 (all quantities are in arbitrary units except for the velocities
that are in km\,s$^{-1}$ with $V_0=2$ km\,s$^{-1}$ and $V_1=8$
km\,s$^{-1}$). They simply scale as $1/R$, reflecting the $1/r^2$
dependence of the effective density. In such a simple model, the flux
ratio between the CO(1-0) and CO(2-1) is a constant, independent of
$R$ and $V_x$.  

\section{Comparison of the observations with a bipolar 
outflow model}\label{compsec}

\begin{figure*}
\centering
\begin{minipage}[c][][t]{1.\textwidth}
\centering
\includegraphics[width=12.0cm,trim=0.cm 0.5cm 0.cm 0.cm,clip]{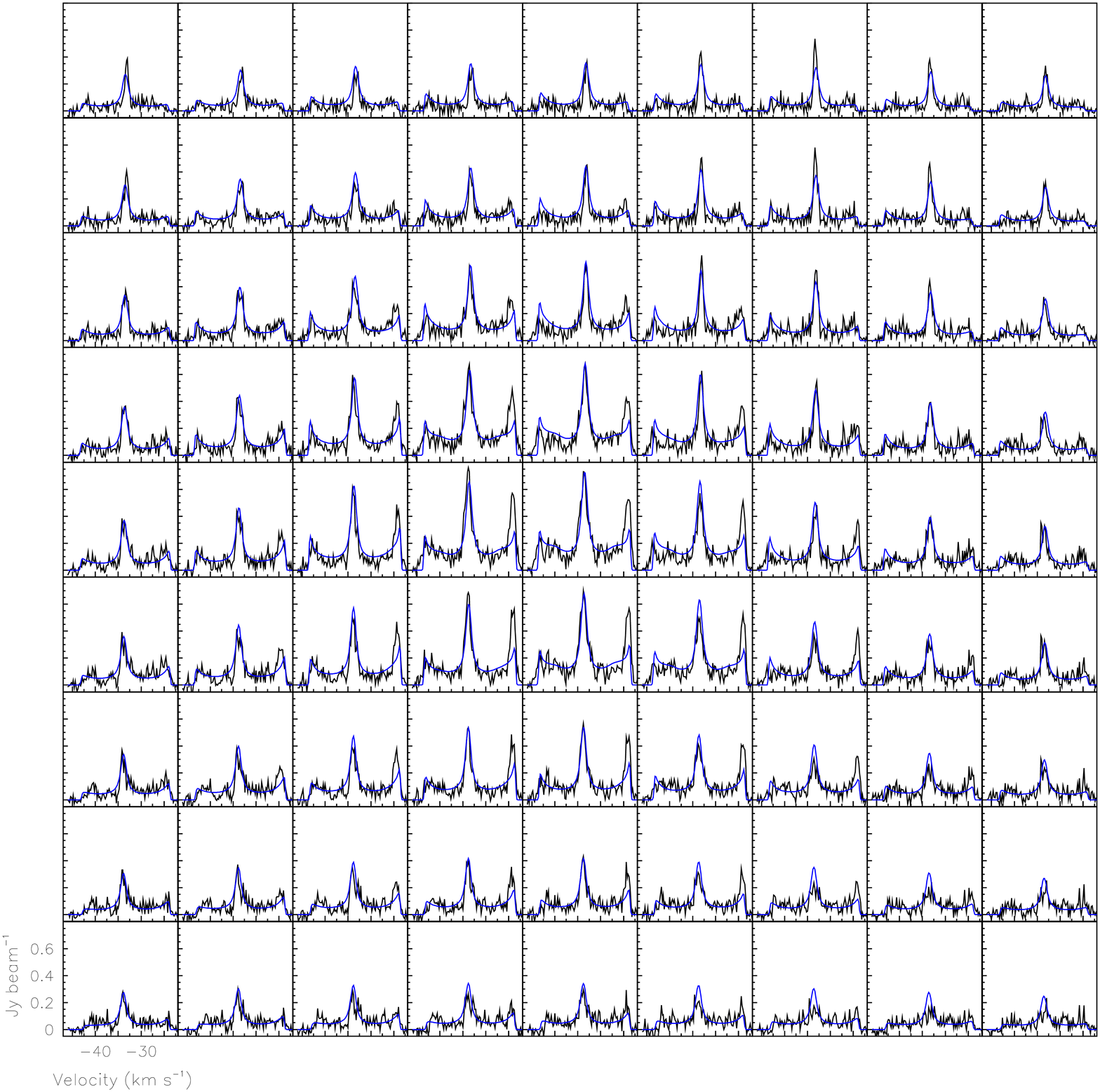}
\includegraphics[width=12.0cm,trim=0.cm 0.5cm 0.cm 1.cm,clip]{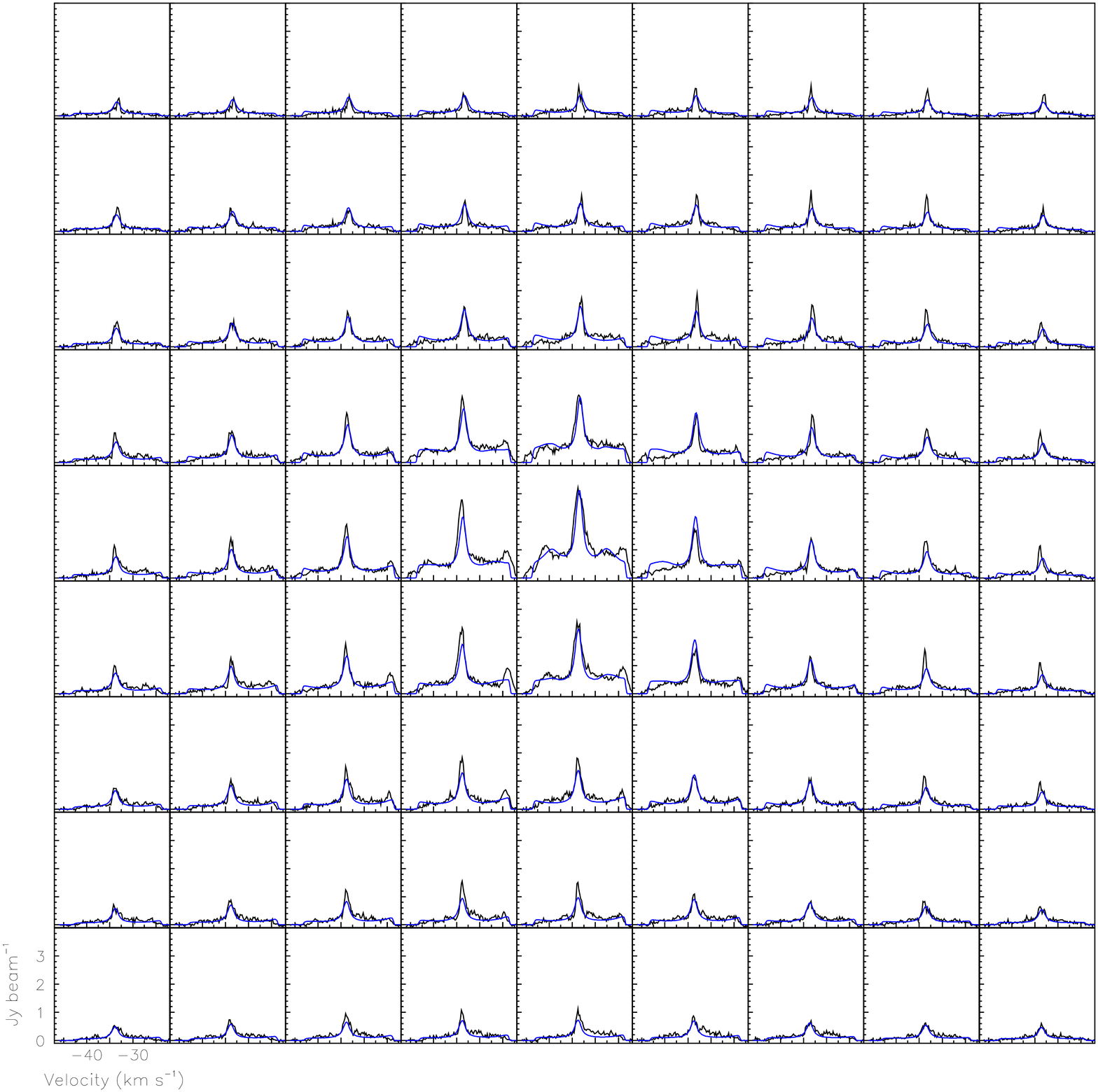}
\caption{Spectral maps centered on the star of the CO observations
(black) and the best-fit model (blue). The CO(1-0) map is shown in the
upper panel, CO(2-1) in the lower panel. Steps in right ascension and
declination are $1''$.}
\label{fig3}
\end{minipage}
\end{figure*}

The spectral maps extracted from the reprocessed data (see
Sect.~\ref{obssec}) are shown in Fig.~\ref{fig3}. The spectra are
displayed in steps of 1$''$ in right ascension and declination. The
synthesized beam is $3.53''\times1.84''$ (PA=28$^\circ$) for CO(1-0),
and $1.67''\times 0.94''$ (PA=29$^\circ$) for CO(2-1), respectively.

To describe EP Aqr, we use the model of Hoai et al.~(2014),
which assumes that the wind is purely radial,  free of turbulence, and
in local thermal equilibrium. For solving the radiative transfer, a
ray-tracing method that takes absorption into account is used (Hoai 2015).
Moreover, the wind is assumed to be stationary and is supposed to have
been in such a  regime for long enough a time, such that the radial
extension of the gas volume is governed exclusively by the UV
dissociation  of the CO molecules by interstellar radiation (Mamon et
al. 1988) and does not keep any trace of the star history. For EP Aqr
the CO/H abundance ratio is taken as 2.5 $10^{-4}$  (Knapp et
al. 1998), a value representative for an M-type star. The temperature
is parametrized by a power law of $r$, $T=T_0r^{-n}$ but is
independent of the stellar latitude $\alpha$. The spectrum of flux
densities in each pixel is calculated by integration along the line of
sight, the temperature dependent contributions of emission and
absorption being respectively added and subtracted at each step. The
wind velocity $V$ and the flux of matter $f_M$ are smooth functions of
$|\sin\alpha|$ and allowance is made for a radial velocity
gradient described  by parameters $\lambda_1$ at the poles and
$\lambda_2$ at the equator. Assuming the wind to be stationary, the
density $\rho$ is then defined at any point by the relation 
$\dot{M}(\alpha)=4\pi f_M(\alpha)=4\pi r^2 \rho(\alpha,r) V(\alpha,r)=const(\alpha)$. 

The bipolarity of the flow is parametrized as a function of
$\sin\alpha$ using Gaussian forms centered at the poles:

\begin{equation}
G=exp[-\frac{1}{2}(\sin\alpha-1)^2/\sigma^2]+exp[-\frac{1}{2}(\sin\alpha+1)^2/\sigma^2],
\label{bipoleq}
\end{equation}

where $\sigma$ is a parameter to be adjusted. This function is used to
define the $\sin\alpha$ dependence of both the wind velocity and the flux of matter:  
\begin{equation}
V = V_1G(1-\lambda_1 e^{-r/2.5''})+V_2(1-\lambda_2 e^{-r/2.5''}) 
\label{velo2eq}
\end{equation}

and 
\begin{equation}
f_M=f_{M_1}G+f_{M_2}.  
\label{mdoteq}
\end{equation}

The density then results from mass conservation:
\begin{equation}
\rho=r^{-2}(f_{M_1}G+f_{M_2})/[V_1G(1-\lambda_1
e^{-r/2.5''})+V_2(1-\lambda_2 e^{-r/2.5''})].
\label{rhoeq}
\end{equation}

For $r$-independent wind velocities, $\lambda_1=\lambda_2=0$, the
velocity and flux of matter are nearly $V_1+V_2$ and
$f_{M_1}+f_{M_2}$ at the poles and $\varepsilon V_1+V_2$ and
$\varepsilon f_{M_1}+f_{M_2}$ at the equator where $\varepsilon$
is the small positive value taken by $G$ at the equator.  

A small inclination $\theta$ of the star axis with respect to the line
of sight, with position angle $\psi$ with respect to the north, is made 
allowance for. The relations quoted in Sect.~\ref{geometrysec} for $\theta=0$
are modified accordingly. More precisely, the position of the star
frame with respect to the sky plane and line of sight is now defined
by angles $\theta$ and $\psi$ and the polar coordinates in the star
frame are the latitude $\alpha$ and the longitude $\omega$ defined in such 
a way as to conform with the definition given in Sect.~\ref{geometrysec} for
$\theta=0$. The star radial velocity as determined from the CO
line profiles is $-33.5$ km\,s$^{-1}$ in good agreement with the
value of $-34$ km\,s$^{-1}$ derived in Winters et al. (2003). 
The values taken by the parameters are obtained by a standard $\chi^2$ minimization method.
The best fit values of the parameters of the
model to the joint CO(1-0) and CO(2-1) observations are listed in
Table~\ref{table1} and the quality of the fit is illustrated in
Fig.~\ref{fig3} where modeled velocity spectra are
superimposed over observed ones.

\begin{table}
\centering
\caption{Best fit parameters obtained for the CO(1-0) and CO(2-1) data. 
A distance of 114 pc is adopted.}
\begin{tabular}{cc}
\hline
Parameter        &    Best fit value    \\
\hline
$\theta$         &    13$^\circ$          \\
$\psi$           &    144$^\circ$         \\
$\sigma$         &    0.3                \\
$V_1$            &    8.0 km\,s$^{-1}$    \\
$V_2$            &    2.0 km\,s$^{-1}$    \\
$f_{M_1}$      & 1.26 10$^{-8}$ \Msold sr$^{-1}$  \\
$f_{M_2}$      & 0.49 10$^{-8}$ \Msold sr$^{-1}$  \\
$\lambda_1$      &    0.52               \\
$\lambda_2$      &    0.38               \\
$T_0$ ($r=1''$)  &    116 K              \\
$n$              &    0.77               \\
\hline
\end{tabular}\\
\label{table1}
\end{table}

This result shows that it is possible to describe the morphology and
kinematics of the gas envelope of EP Aqr as the combination of a slow
isotropic wind and a bipolar outflow, the wind velocities at the
equator and at the poles being ${\sim}2\,$km\,s$^{-1}$ and
${\sim}10\,$km\,s$^{-1}$ respectively and the inclination of the star
axis with respect to the line of sight, $\theta$, being
${\sim}13^\circ$. The flux of matter varies from 0.49 10$^{-8}$
\Msold sr$^{-1}$ in the equatorial plane to 1.26 10$^{-8}$ \Msold
sr$^{-1}$ in the polar directions. The total mass loss rate (integral of 
$\dot{M}(\alpha)$ over $\alpha$) is 1.8
10$^{-7}$ \Msold. The angular aperture of the bipolar outflow,
measured by the parameter $\sigma = 0.3$ in Eq.~(\ref{bipoleq}),
corresponds to  $\alpha=\arcsin(0.7)\sim 45^\circ$.

Leaving for Sect.~\ref{narrowsec} a more detailed discussion of
this result, we study in the next section the CO(1-0) to CO(2-1) flux
ratio, which gives important information on the temperature
distribution with only minimal dependence on the details of the
model. 

\section{The CO(1-0) to CO(2-1) flux ratio}\label{fluxsec}
\begin{figure}
\centering
\includegraphics[width=9.cm,trim=0.cm 0.5cm 0.cm 0.5cm,clip]{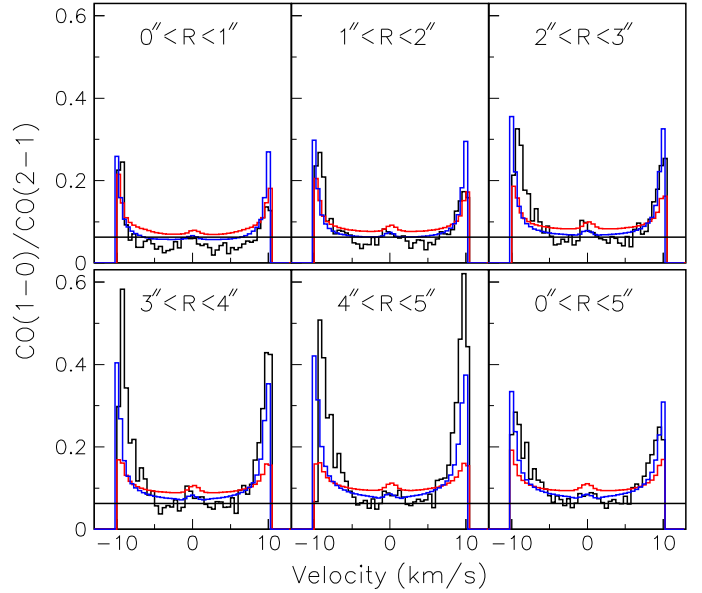}
\caption{CO(1-0) to CO(2-1) flux ratio (black) where each of the 
\mbox{CO(1-0)} and CO(2-1) fluxes has been averaged over 5 successive velocity 
bins and over the concentric rings defined in the text. The lower right panel 
is for all pixels having $R<5''$. The result of the best fit of 
the model described in 
Sect.~\ref{compsec} is shown in red and that of its modification described in 
Sect.~\ref{fluxsec} is shown in blue. The horizontal lines indicate the level 
(0.063) above which the data must be under the hypothesis of local thermal 
equilibrium.}
\label{fig4}
\end{figure}

Comparing the fluxes associated with CO(1-0) and CO(2-1) emission 
provides information on the gas temperature nearly independently from
the gas density and velocity, which are common to the two lines. In
the optically thin limit and assuming local thermal equilibrium, the
temperature $T$(K) is obtained from the ratio $Q$ of the CO(1-0) to
CO(2-1) fluxes as $T=11.1/\ln(Q/0.063)$. However, the model presented
in the preceding section does not reproduce well the observed flux
ratio. In particular, we observe that $Q$ is sometimes below
the minimal value 0.063 allowed by the local thermal equilibrium
hypothesis (when $T=\infty$). This is illustrated in Fig.~\ref{fig4},
where the data have been averaged over concentric rings limited by
circles having $R$=1$''$, 2$''$, 3$''$, 4$''$ and 5$''$ respectively. 

\begin{figure}
\centering
\includegraphics[height=4.5cm,trim=1.5cm .5cm 2.cm 2.cm,clip]{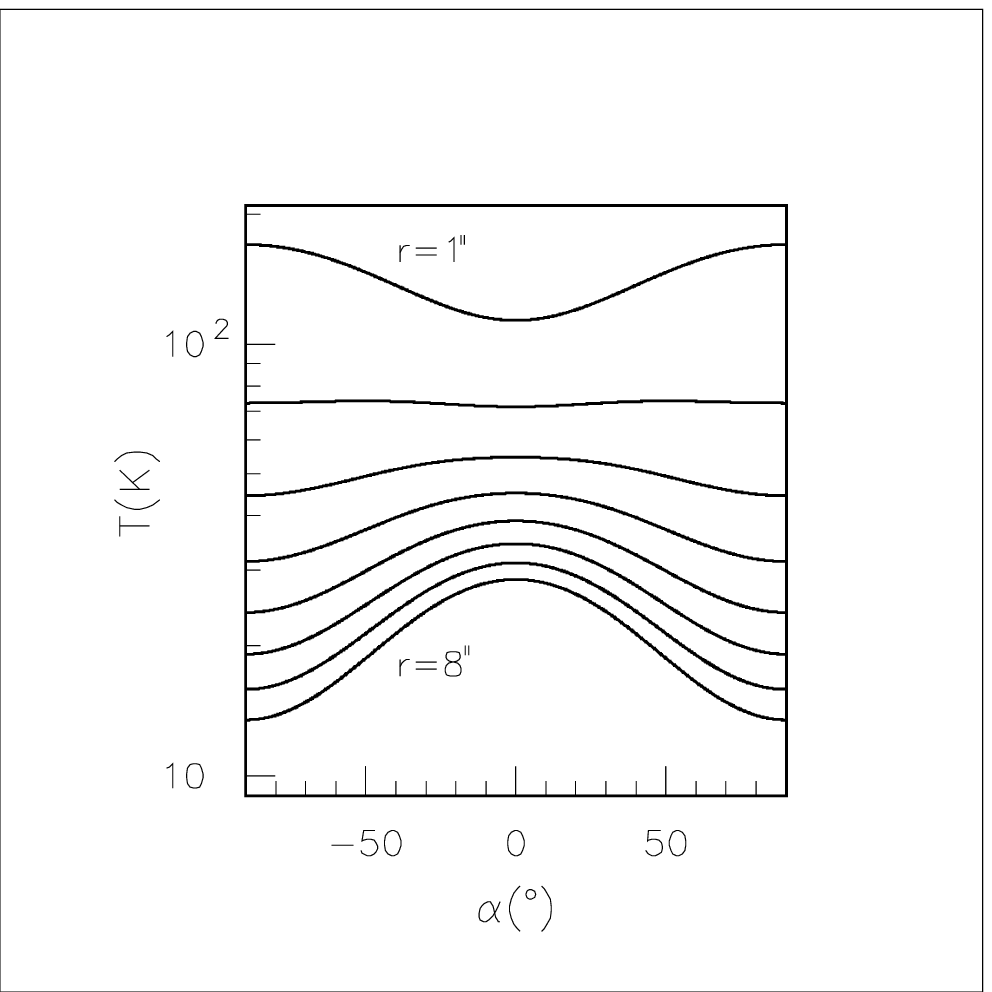}
\includegraphics[height=4.5cm,trim=0.5cm .5cm 2.cm 2.cm,clip]{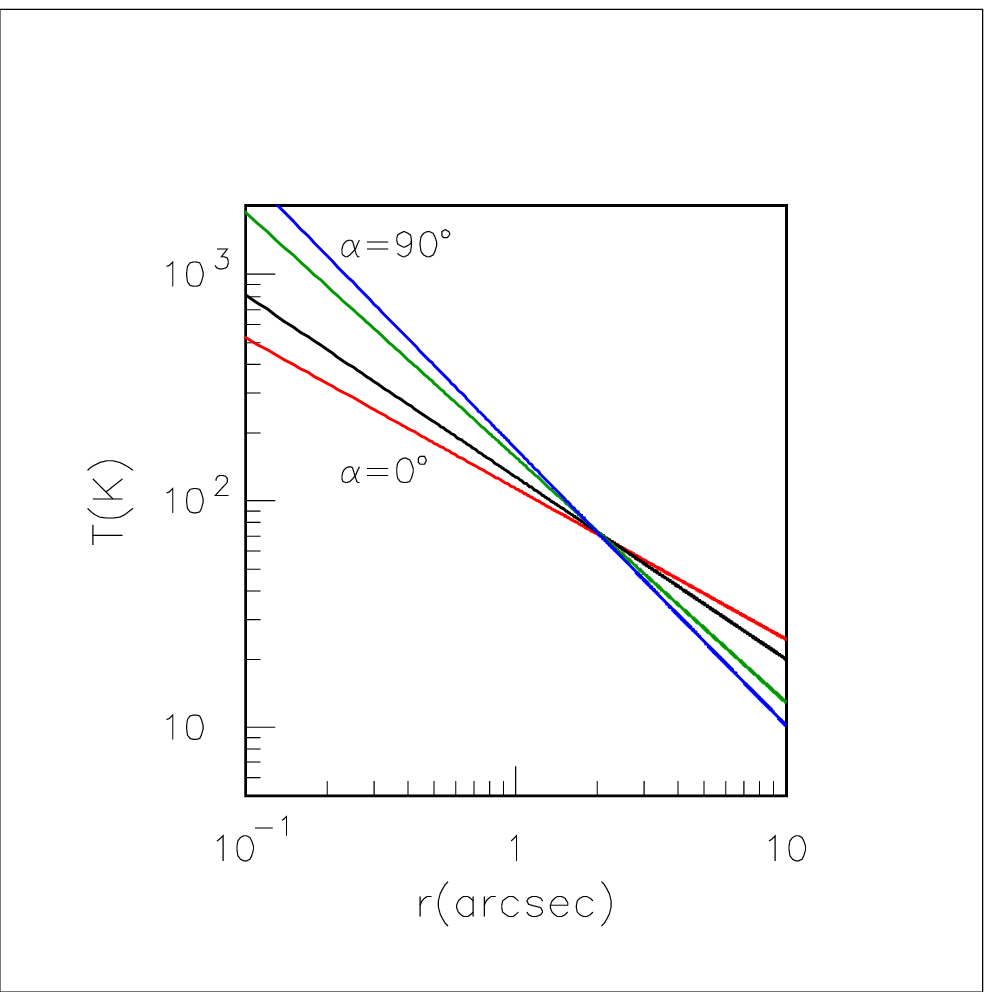}
\caption{Distribution of the gas temperature. Left: as a function of
$\alpha$ at distances from the star $r= 1''$ to $8''$ (from top to
bottom) in steps of $1''$; right: as a function of $r$ at latitudes
$\alpha=0^\circ$ (red), $30^\circ$ (black), $60^\circ$ (green) and
$90^\circ$ (blue).}
\label{fig5}
\end{figure}

Accordingly, we allow for a small difference of normalization between
the CO(1-0) and CO(2-1) fluxes by multiplying the former by a factor
$1+\mu$ and the latter by a factor $1-\mu$. This factor may account
for a real difference in calibration as well as, in an {\it ad hoc}
way, for any inadequacy of the model (assumption of local thermal
equilibrium, absence of turbulence). 
Moreover, we see from Fig.~\ref{fig4} that one needs to
increase the ratio $Q$ near the extremities of the velocity spectrum
more than in the middle, meaning near the poles at large values of $r$
more than near the equator at small values of $r$. Accordingly, we
parametrize the temperature in the form
\begin{equation}
T=T_0(\sin^2\!\alpha+a\cos^2\!\alpha)(r/1'')^{-n}
\label{tempeq}
\end{equation} 
with
\begin{equation}
n=n_1+n_2\cos^2\!\alpha.
\label{tindexeq}
\end{equation} 
At $r=1''$, the temperature is $T_0$ at the
poles and $aT_0$ at the equator. The power index describing the radial
dependence of the temperature is $n_1$ at the poles and $n_1+n_2$ at
the equator. 

The result is summarized in Table~\ref{table2} and illustrated in
Fig.~\ref{fig5}. The ratio between the values taken by the temperature
at the equator and at the poles, $ar^{-n_2}$, varies from 0.70 at
$r=1''$ to 1.64 at $r=5''$, crossing unity at $r\approx2''$.  As $r/R$
is much larger near the poles than at the equator (it is equal to
1/cos$\alpha$ in the simple configuration studied in
Sect.~\ref{geometrysec}), the extremities of the velocity spectrum
probe large  $r$ values, meaning low temperatures, as required by the
measured $Q$ values: both the values taken by $n_2$ and by $a$ are
determinant in increasing $Q$ at the extremities of the velocity
spectrum. For the same reason, the temperature increase obtained from
the parametrization at small values of $r$ and large values of
$\alpha$ is not probed by any of the data, which makes its reliability
uncertain. 

The values of the temperature that we find close to the central star 
($\sim$ 1000 K at 0.1$''$) are consistent with those obtained by Cami et al. 
(2000) using infrared CO$_2$ lines.

\begin{table}
\centering
\caption{Best fit parameters of the CO(2-1) to CO(1-0) ratio.}

\begin{tabular}{ccccccc}
\hline
\multicolumn{1}{c}{Parameter}       & $T_0$(K) &   a  & $n_1$ & $n_2$   & $\mu$ (\%) & $\chi^2/dof$ \\
\hline
\multicolumn{1}{c}{Best fit values} &    170   & 0.67 & 1.22  & $-$0.55 &      9.0   & \multicolumn{1}{c}{\multirow{2}{*}{1.14}}\\
                     Uncertainties  &     20   & 0.12 & 0.06  &  0.10   &      3.0   & \multicolumn{1}{c}{\multirow{2}{*}{}} \\
\hline  
\end{tabular}
\label{table2}
\end{table}

The uncertainties listed in Table~\ref{table2} are evaluated from the
dependence of $\chi^2$ over the values of the parameters, taking
proper account of the correlations between them. However, they have
been scaled up by a common factor such that the uncertainty on $T_0$
be 20 K. This somewhat arbitrary scaling is meant to cope with our
lack of detailed control over systematic errors and the value of 20 K
is evaluated from the robustness of the results against changes of
different nature that have been made to the model finally adopted in
the course of the study. While the quoted uncertainties give a good
idea of the reliability of the result and of its sensitivity to the
values taken by the parameters, both the quality of the data (in
particular the need for a $\mu$-correction) and the crudeness of the
model do not allow for a more serious quantitative evaluation.

We noted that the quality of the fits can be slightly improved by
allowing for a temperature enhancement at mid-latitudes and small
distances to the star; however, we were unable to establish the
significance of this result with sufficient confidence.

The following two conclusions can then be retained:
\begin{itemize}
 \item At large distances to the star, the temperature is
significantly higher at the equator than at the poles, by typically 15
K at $r=5''$;
 \item The temperature decreases with distance as ${\sim} r^{-1.2}$ at
the poles and significantly less steeply at the equator, typically as
$r^{-0.7}$. 
\end{itemize}

\section{Evaluation of the effective density in the star meridian plane} \label{meridplanesec}

Under the hypotheses used to evaluate the model described in
Sect.~\ref{compsec}, one can associate to each data-triple
$(y,z,V_x)$ a point  $(r,\alpha)$ in the meridian plane of the
star. Indeed, $x$ is a known function of $r$, $\alpha$, and the star
longitude $\omega$, as is $V_x=Vx/r$. The former uses the values
of $\theta$ and $\psi$ and the latter the parametrization of $V$ as a
function of $r$ and $\alpha$ that were obtained from the best fit. As
one has three equations and three unknowns one obtains $r$, $\alpha$
and $\omega$ given $V_x$, $y$ and $z$. The flux density $f$ 
corresponding to a given data-triple can therefore be mapped as
an effective density $\rho_\mathrm{eff}=f\mathrm{d}V_x/\mathrm{d}x$ in
the meridian plane of the star, (i.e. each plane that contains
the star's polar axis) with coordinates $r\cos\,\alpha$ and
$r\sin\,\alpha$ (see Fig.~\ref{fig1}). Its evaluation uses the
best fit parametrization of the wind velocity but not that of the flux
of matter: it provides therefore a consistency check.

\begin{figure}
\centering
\includegraphics[width=4.3cm,height=6.5cm,trim=0.5cm 0.5cm 0.cm 0.8cm,clip]{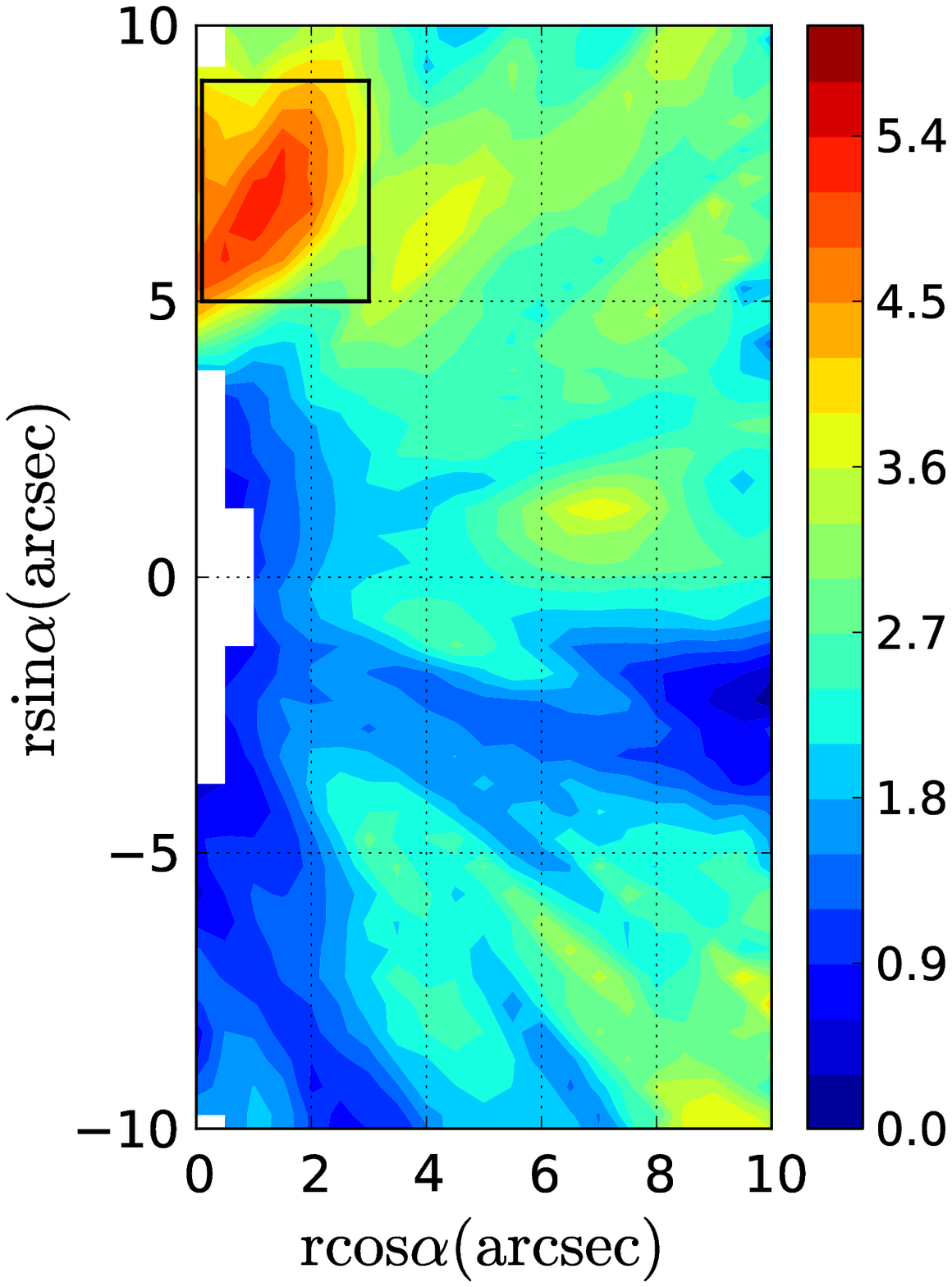}
\includegraphics[width=4.3cm,height=6.5cm,trim=0.5cm 0.5cm 0.5cm 0.8cm,clip]{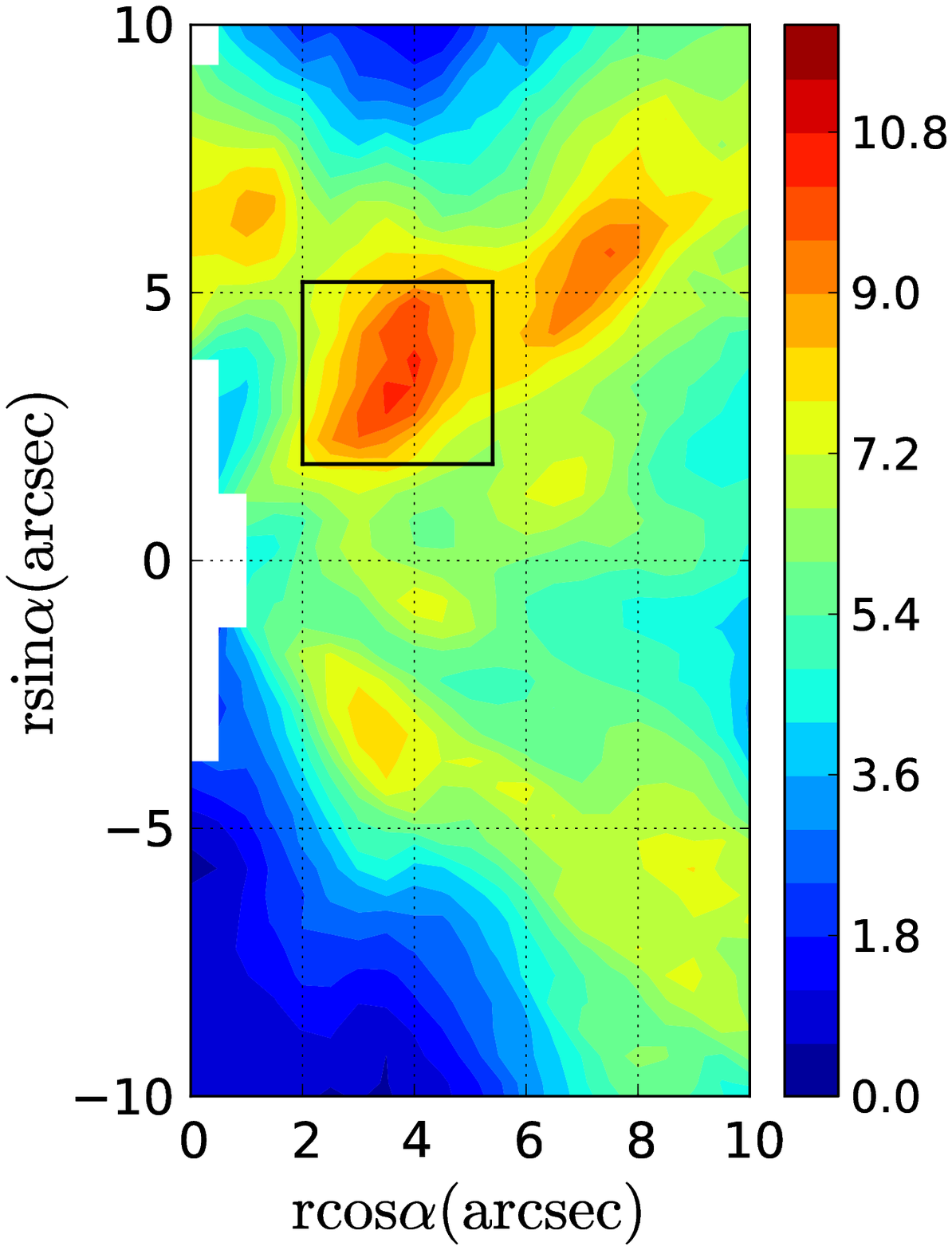}
\caption{Reconstructed maps of the effective density, multiplied by
$r^2$, in the star meridian plane under the assumption of a wind velocity having
the form obtained from the best fit of the model of
Sect.~\ref{compsec}.  The colour codes are such that the ratio between
maximum and minimum values of $\rho r^2$ are the same for CO(1-0)
(left) and CO(2-1) (right). Units are Jy beam$^{-1}$ km\,s$^{-1}$
arcsec. The rectangles show the regions selected for displaying the
$\omega$ distributions shown in Fig.~\ref{fig8} (for CO(1-0), abscissa
between $0''$ and $3''$, ordinate between $5''$ and $9''$; for
CO(2-1), abscissa between $2.0''$ and $5.4''$, ordinate between
$1.8''$ and $5.2''$). }
\label{fig6}
\end{figure}

\begin{figure}
\centering
\includegraphics[width=8.cm,trim=1.5cm 1.cm .8cm 9.cm,clip]{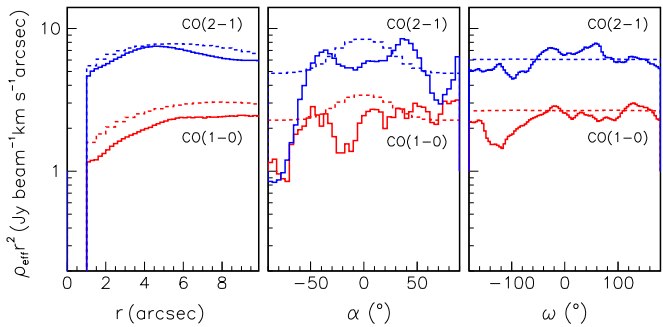}
\caption{Distributions of the average effective density, multiplied by
$r^2$, reconstructed in the meridian plane of the star for a wind
velocity having the form obtained from the best fit of the model of
Sect.~\ref{compsec}.  Note that the $\mu$-correction has not been
applied.  left: radial dependence of $r^2\rho_\mathrm{eff}$,
averaged over $\alpha$ and $\omega$; middle: latitude dependence of
$r^2\rho_\mathrm{eff}$, averaged over $r$ and $\omega$; right:
longitude  dependence of $r^2\rho_\mathrm{eff}$, averaged over $r$ and
$\alpha$. In each panel, CO(1-0) results are shown in red and
CO(2-1) results in blue. The dashed curves show the results of the
model, ignoring absorption. }
\label{fig7}
\end{figure}

The validity of the procedure has been checked by replacing the data
by the results of the model and verifying that the
reconstructed effective density is identical to that used as
input. The region of large $|\alpha|$ values is probed exclusively by
the central pixels near the extremity of the spectrum and is therefore
subject to large systematic uncertainties. Indeed, large $|\alpha|$
values mean large $|V_x|$ values, which in turn mean large $r/R$
values, namely $R$ much smaller than $r$. For this reason, on the
basis of the result of the validity check, we restrict the space
reconstruction to pixels having $R>1''$. In addition, in order to
avoid dealing with too low fluxes, we require $R$ not to exceed
10$''$. The resulting maps of $r^2\rho_\mathrm{eff}$ are displayed in
Fig.~\ref{fig6} for CO(1-0) and CO(2-1) separately and its averaged
distributions as functions of $r$, $\alpha$ and $\omega$ in
Fig.~\ref{fig7}. 

\begin{figure}
\centering
\includegraphics[width=6.cm,trim=1.cm 1.cm 1.cm 0.5cm,clip]{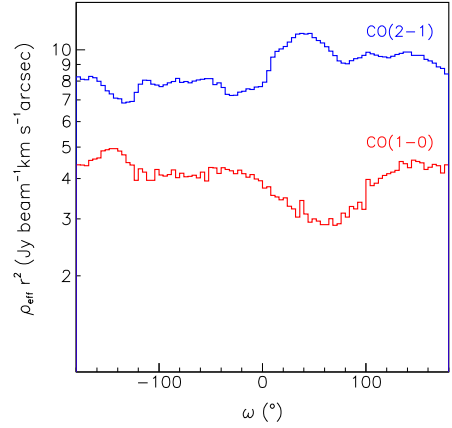}
\caption{Distributions of the effective density, multiplied by $r^2$,
as a function of star longitude $\omega$ for the annular regions
delineated by the rectangles shown in Fig.~\ref{fig6}. CO(1-0) data
are shown in red, CO(2-1) data in blue.}
\label{fig8}
\end{figure}

All distributions display significant deviations with respect to the
distributions generated by the model, which, however, reproduces well
the general trends. In principle, one should be able to include in the
model the result of the space reconstruction of $\rho_\mathrm{eff}$
and iterate the fitting procedure with a more realistic
parametrization of the flux of matter and possibly a more
sophisticated velocity model. However, the quality of the present data
is insufficient to undertake such an ambitious programme. We shall be
satisfied with simply taking note of the differences between the model
of Sect.~\ref{compsec} and the image  revealed by the space
reconstruction, as long as they can be considered as being both
reliable and significant. 

\begin{figure*}
\centering
\includegraphics[width=14.5cm,trim=1.cm 0.cm 0.cm 0.5cm,clip]{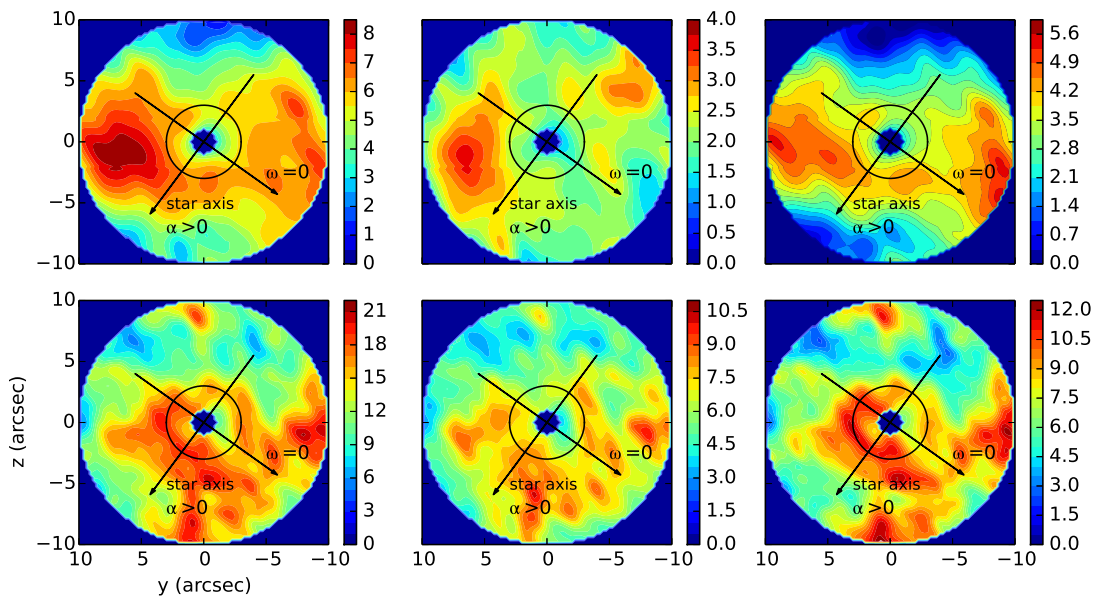}
\caption{Sky maps of $Rf$ for CO(1-0) (upper panels) and CO(2-1)
(lower panels) observations. Units are Jy beam$^{-1}$ km\,s$^{-1}$
arcsec. From left to right: all velocities, $|V_x|<2$ km\,s$^{-1}$ and
$|V_x|>2$ km\,s$^{-1}$. The circles at $R=3.5''$ correspond to the
enhancement seen by Winters et al. (2007) in CO(2-1) data restricted
to the narrow velocity component. The projection of the star axis on
the sky plane (making an angle of $13^\circ$ with the line of sight)
and the axis from which the star longitude $\omega$ is measured
(positive clockwise) are shown as black arrows.}
\label{fig9}
\end{figure*}

The maps of Fig.~\ref{fig6} and the $\alpha$ distributions of
Fig.~\ref{fig7} give evidence for a significant asymmetry between the
northern and southern hemispheres of the star (defining north and
south in the star coordinate system, north corresponding to
$\alpha=\pi/2$, away from Earth, with a projection  towards North
in the sky plane). Indeed, a modification of the model allowing for
an asymmetry of the same form as used in Nhung et al. (2015) for RS
Cnc, namely multiplying the flux of matter by a common factor
$1+\gamma\sin\alpha$, gives a 20\% excess in the northern hemisphere
of the star. Differentiating between the asymmetries of the two mass
loss terms, $f_{M_1}G$ and $f_{M_2}$ reveals a strong correlation
between them, the asymmetry of the first term being more efficient at
improving the quality of the fit. Restricting the asymmetry to the
first term results in a 37\% excess in the northern
hemisphere. The effect of absorption does not exceed ${\sim}5$\% on
average and cannot be invoked to explain such asymmetry. The
fluctuations observed in the distributions of the effective density,
as well as the differences observed between the \mbox{CO(1-0)} and
\mbox{CO(2-1)} data, make it difficult to locate precisely the
northern excess in space. Fig.~\ref{fig8} displays the $\omega$
distributions of the effective density averaged over the $r$-$\alpha$
rectangles drawn in Fig.~\ref{fig6} around the maxima of the meridian
maps; they do not show an obvious enhancement, the maximum observed in
CO(2-1) corresponding in fact to a minimum in CO(1-0). This suggests
that the northern excess is distributed over the whole range of star
longitudes rather than being confined to a particular region in
space. This is confirmed by the $(y,z)$ sky maps of the measured
fluxes (integrated over Doppler velocity and multiplied by $R$) that
are displayed in Fig.~\ref{fig9} and show an enhancement at radii
exceeding ${\sim}3''$. This enhancement, which was already noted by
Winters et al. (2007) in the case of the narrow velocity component for
CO(2-1), is also visible in the broad velocity component. A better
space resolution would be necessary to make more quantitative
statements concerning this excess, its precise location, morphology
and velocity distribution. The present data allow for a
qualitative description, retaining as likely the presence of a
northern enhancement at distances from the star exceeding $3''$ and
distributed more or less uniformly in star longitude. 

\section{The mean Doppler velocity of the narrow line component}\label{narrowsec}

\begin{figure}
\centering
\includegraphics[width=5.5 cm,trim=2.cm 1.cm 5.cm 3.cm,clip]{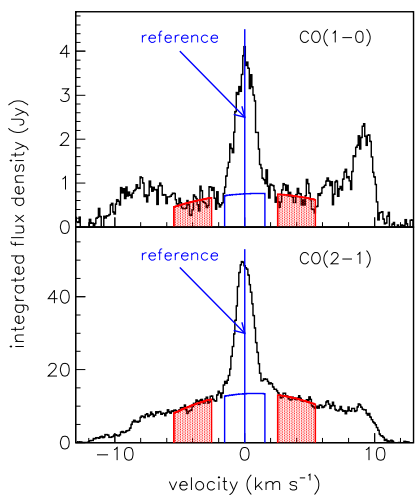}
\caption{Illustration of the procedure used to evaluate the mean
Doppler velocity $\Delta v$ of the narrow component. The velocity
spectra summed over the 1369 pixels of the map are shown in black for
CO(1-0) (upper panel) and CO(2-1) (lower panel). The quadratic fit
over the two control regions is shown in red and its interpolation
below the narrow component in blue. The vertical blue lines show the
mean Doppler velocities of the narrow component from which the
interpolated broad component has been subtracted. They are used as
reference for evaluating $\Delta v$ in each pixel separately.}
\label{fig10}
\end{figure}

\begin{figure}
\centering
\includegraphics[height=4.4 cm,trim=1.cm 0.5cm 0.cm 0.5cm,clip]{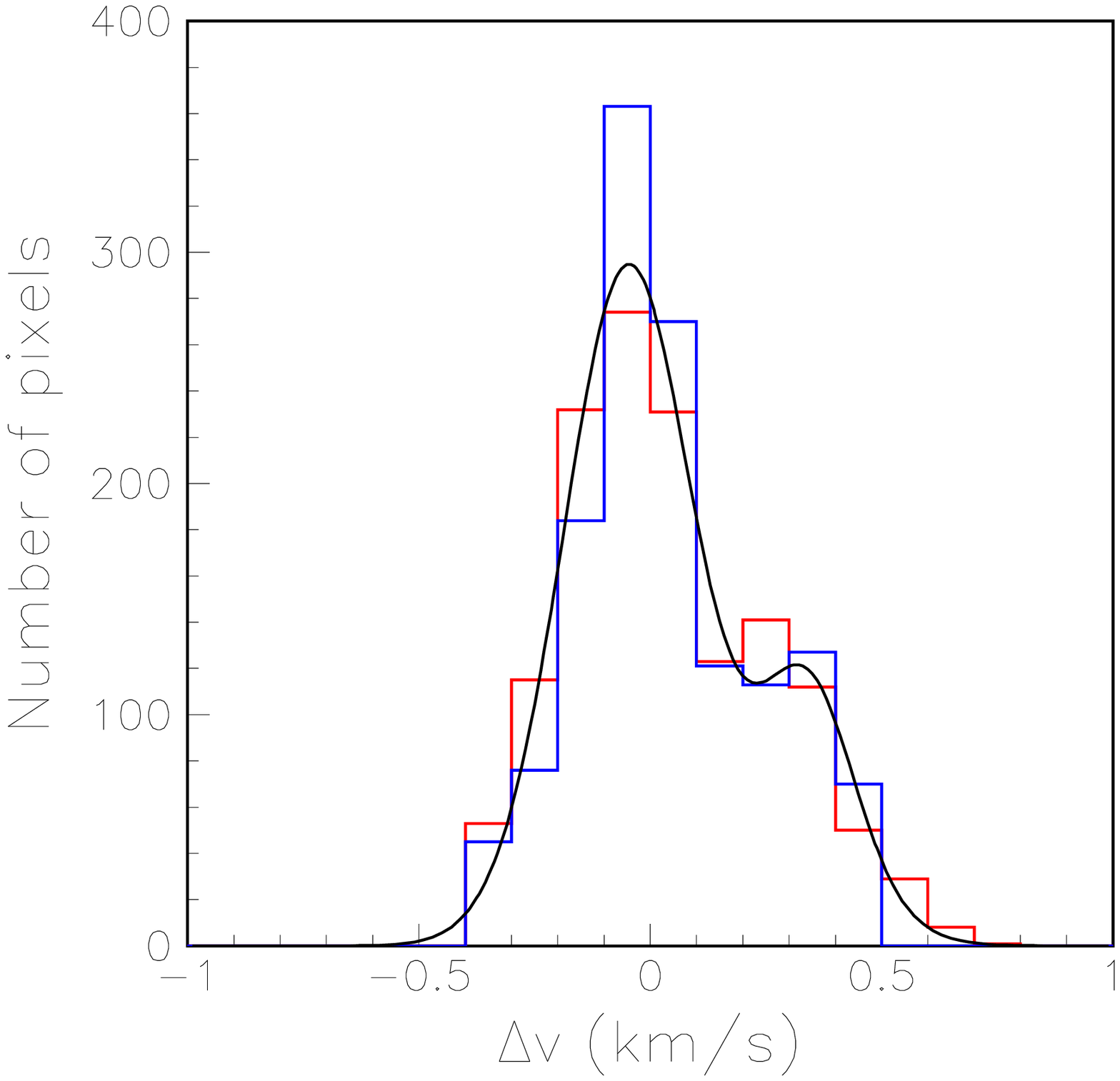}
\includegraphics[height=4.4 cm,trim=1.cm 0.5cm 1.2cm 0.5cm,clip]{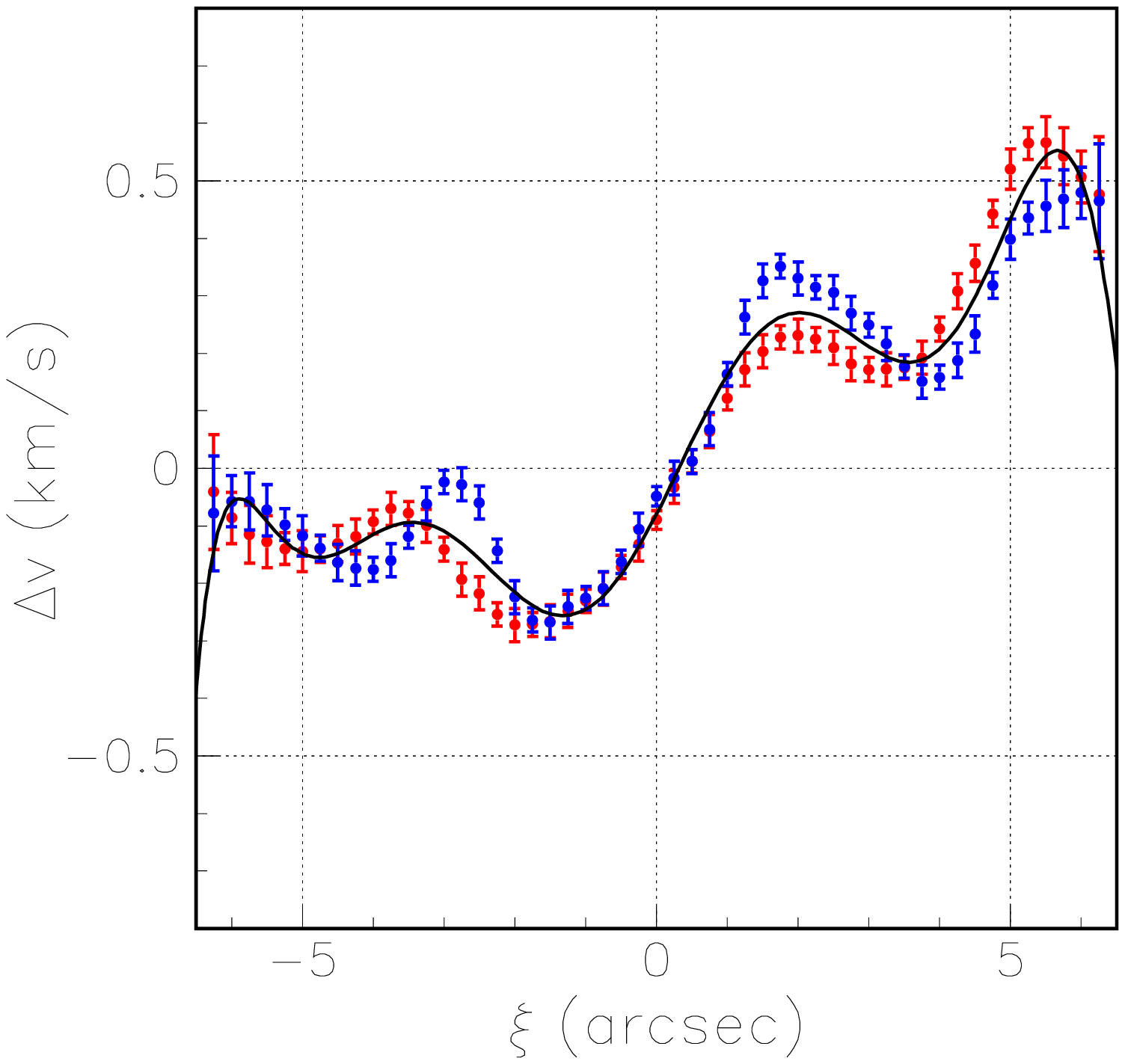}
\caption{Left: distributions of the mean Doppler velocity $\Delta v$
(km\,s$^{-1}$) measured with respect to its value averaged over the
whole map (shown as "reference" in Fig.~\ref{fig10}). The black curve
shows a two-Gaussian common fit to the two distributions. Right:
dependence of the projection of the mean Doppler velocity $\Delta v$
(km\,s$^{-1}$), averaged over pixels included in the bands shown in
Fig.~\ref{fig12}, on coordinate $\xi$ measured from south-east to
north-west. The black curve shows a polynomial common fit to the two
distributions. In both panels the CO(1-0) data are shown in red and
the CO(2-1) data in blue.}
\label{fig11}
\end{figure}

\begin{figure*}
\centering
\includegraphics[width=6.cm,trim=1.cm 0.5cm 1.cm 0.5cm,clip]{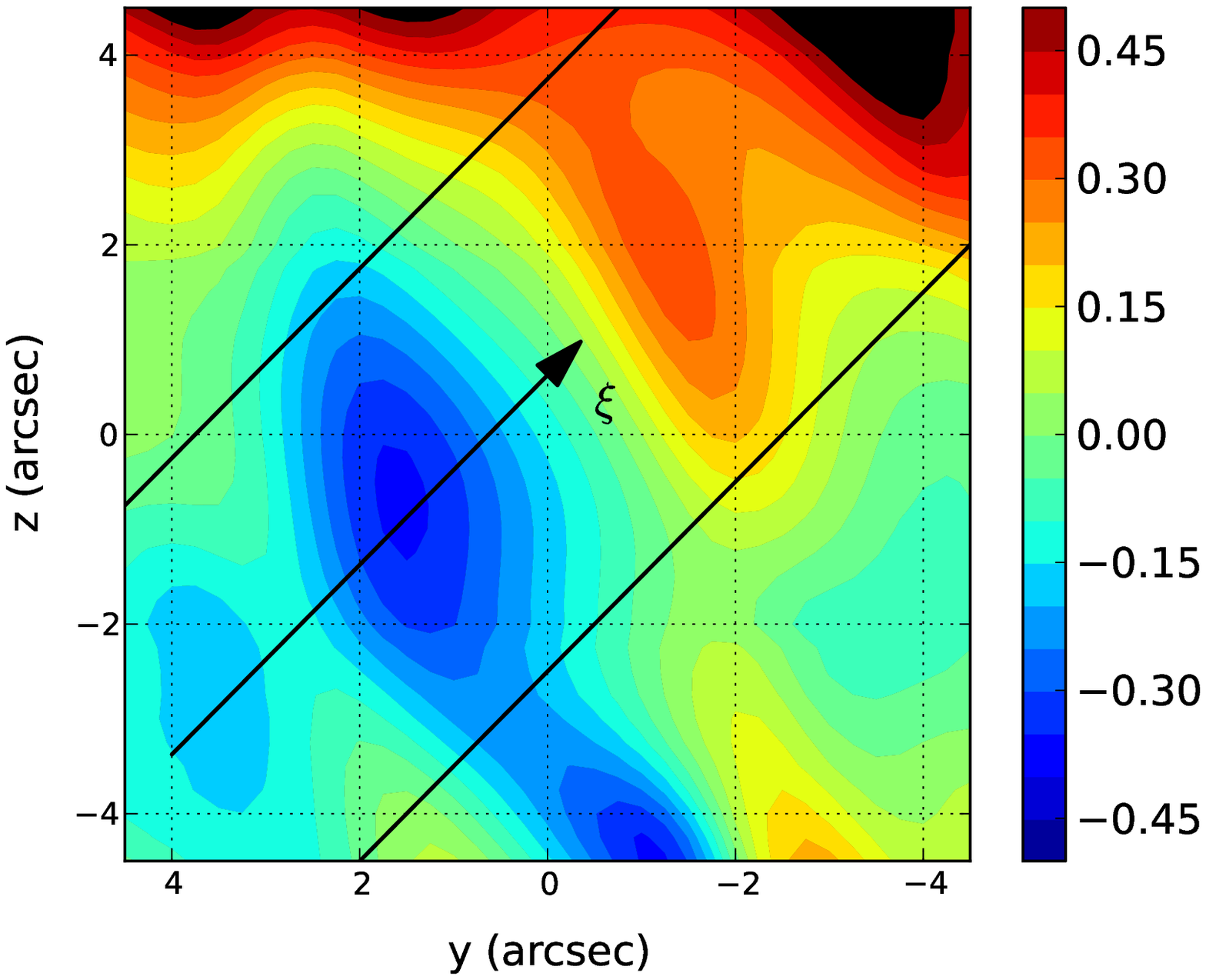}
\includegraphics[width=6.cm,trim=1.cm 0.5cm 1.cm 0.5cm,clip]{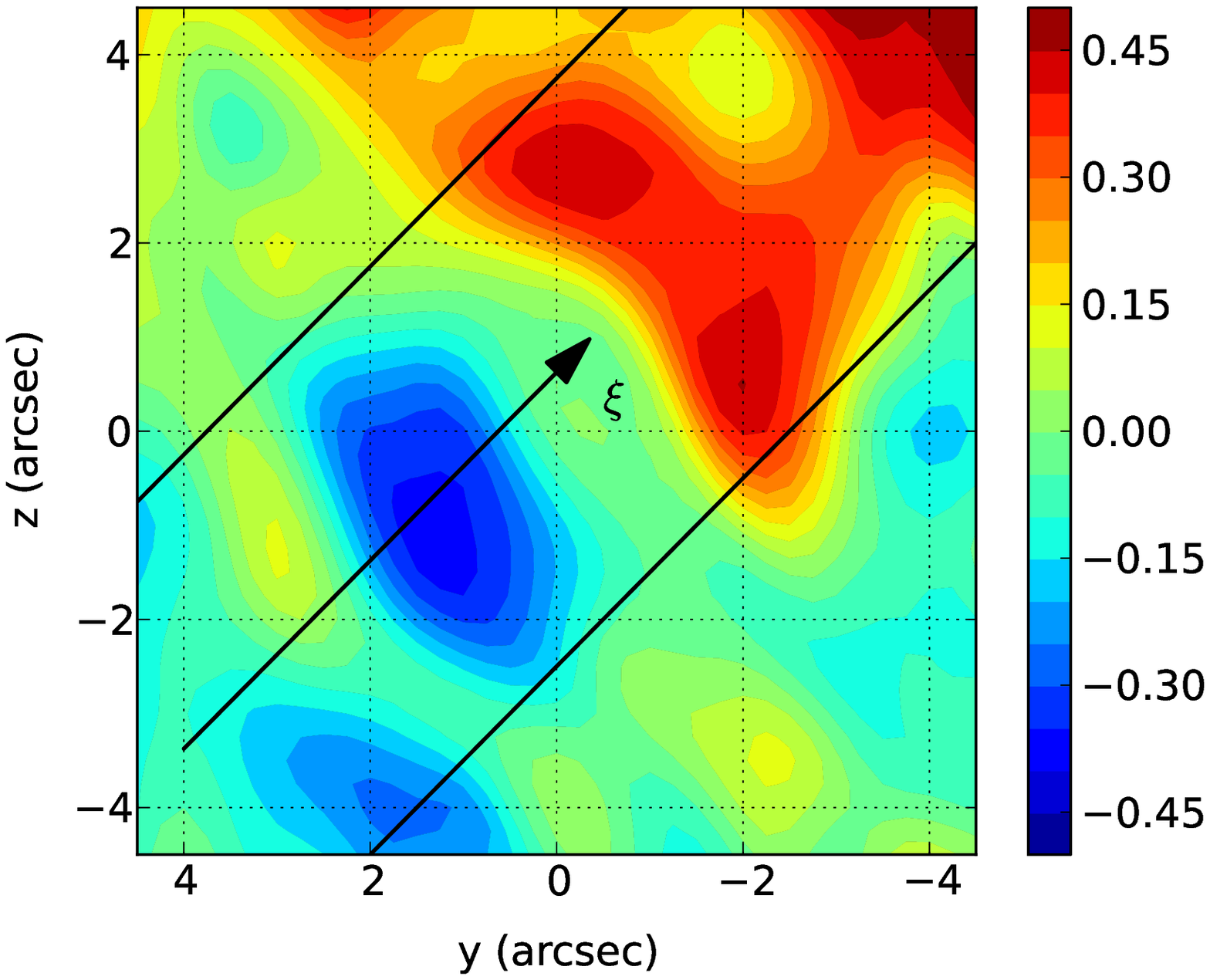}
\caption{Sky maps of the mean Doppler velocity $\Delta v$
(km\,s$^{-1}$) measured with respect to its value averaged over the
whole map (shown as "reference" in Fig.~\ref{fig10}) for CO(1-0)
(left) and CO(2-1) (right). The black lines limit the bands in which
pixels are retained to evaluate the $\xi$ dependence of $\Delta v$
(Fig.~\ref{fig11} right).}
\label{fig12}
\end{figure*}

The centroid of the narrow component of the observed spectra
moves across the sky map by ${\sim}\pm0.5$ km\,s$^{-1}$ on either side
of the average velocity. We measure its mean velocity $\Delta v$ in
each pixel after subtraction of the underlying broad component,
interpolated from control regions spanning velocity intervals
between 5.4\,km\,s$^{-1}$ and 2.6\,km\,s$^{-1}$ with respect to the
average velocity (either blue-shifted or red-shifted). The procedure
is illustrated in Fig.~\ref{fig10} for the spectra summed over
the 37$\times$37=1369 pixels of the sky map
$(9.25''\times9.25'')$. The resulting $\Delta v$ distributions,
measured in km\,s$^{-1}$ with respect to the mean velocity
measured for the whole maps, are shown in Fig.~\ref{fig11} (left) and
their maps in Fig.~\ref{fig12}. Both CO(1-0) and CO(2-1) data display
very similar features, the narrow component being seen to be
red-shifted in the north-west direction and blue-shifted in the
south-east direction, nearly parallel to the direction of the
projected star axis. This suggests that the inclination of the star
axis with respect to the line of sight, measured by the angle
$\theta{\sim}13^\circ$, might be the cause of the effect.

\begin{figure}
\centering
\includegraphics[height=4.4 cm,trim=1.cm 0.5cm 0.cm 0.5cm,clip]{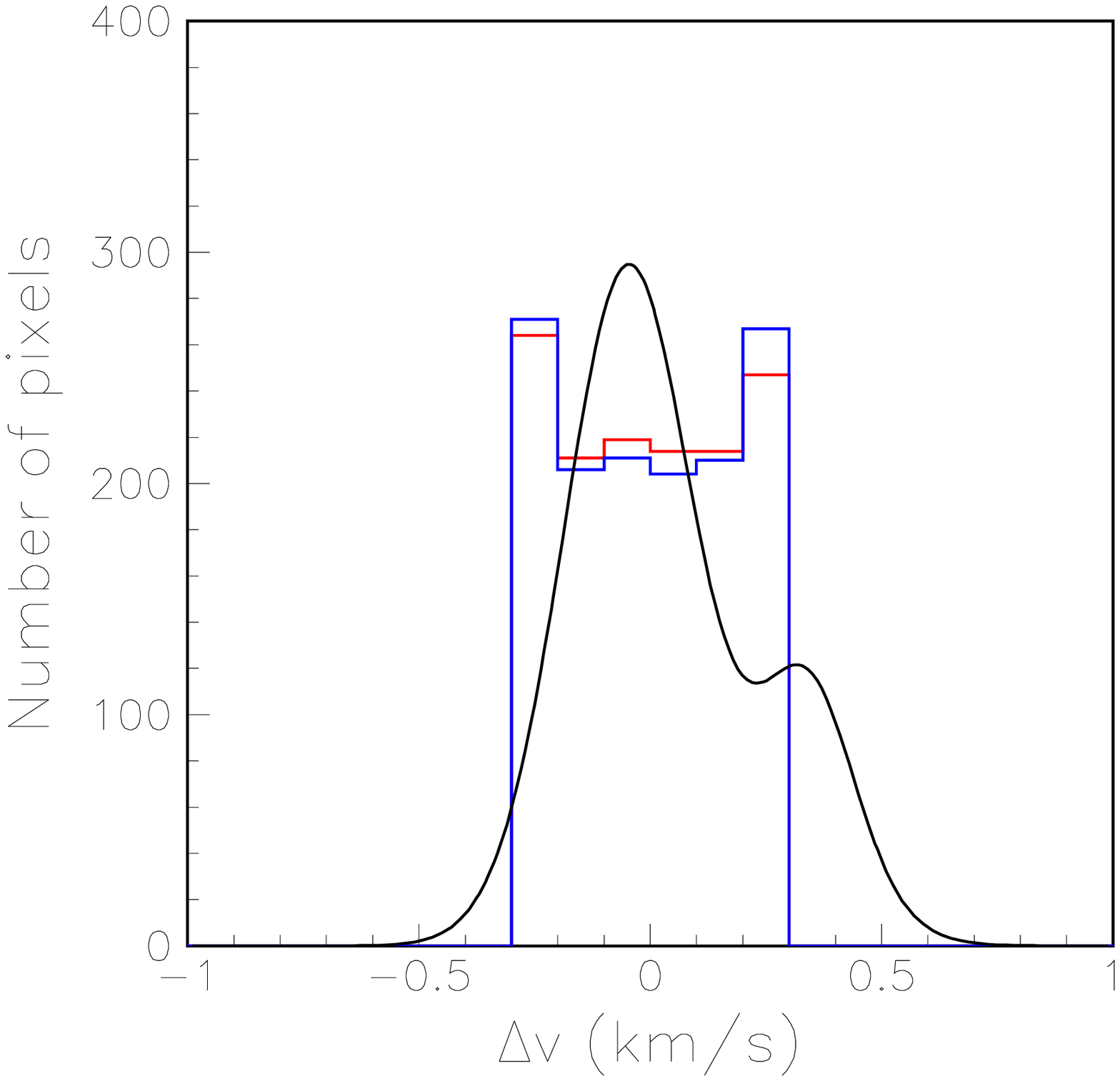}
\includegraphics[height=4.4 cm,trim=1.cm 0.5cm 1.2cm 0.5cm,clip]{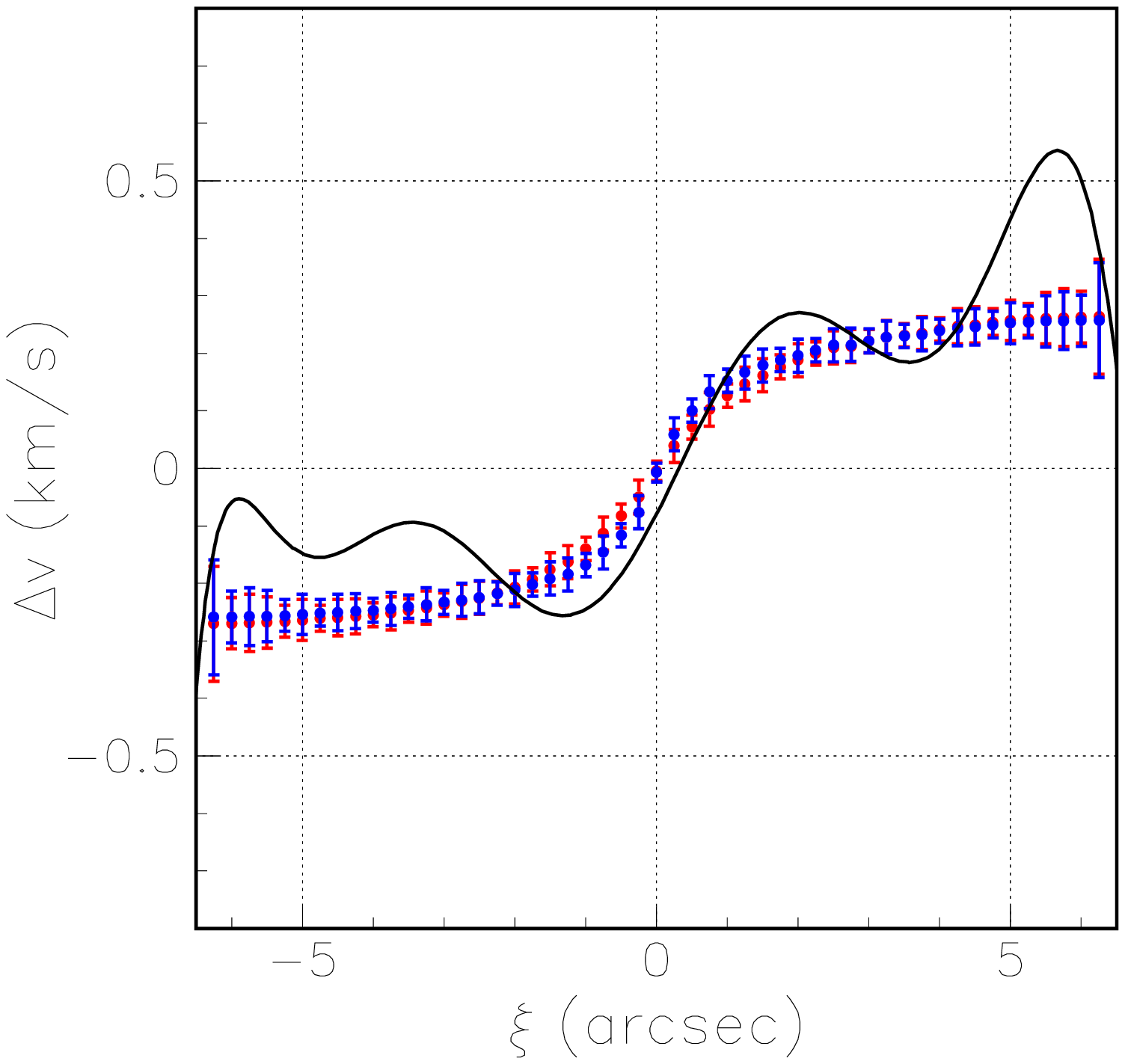}
\caption{Results of the model introduced in Sect.~\ref{compsec}
for the distributions displayed in Fig.~\ref{fig11}. The smooth curves
are the results of the fits (respectively Gaussian and polynomial)
made to the observations in Fig.~\ref{fig11}.}
\label{fig13}
\end{figure}

\begin{figure*}
\centering
\includegraphics[width=6.cm,trim=1.cm 0.5cm 1.cm 0.5cm,clip]{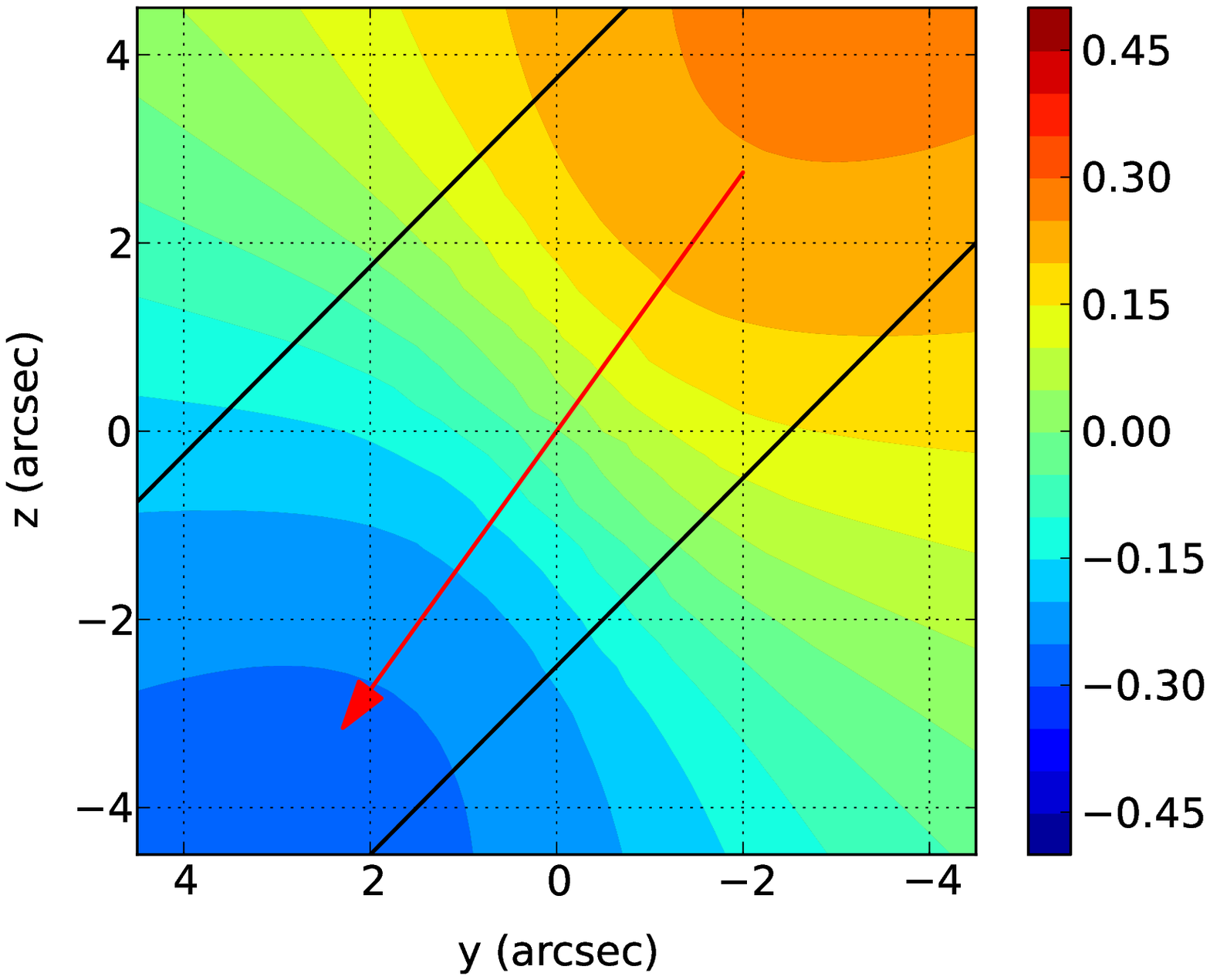}
\includegraphics[width=6.cm,trim=1.cm 0.5cm 1.cm 0.5cm,clip]{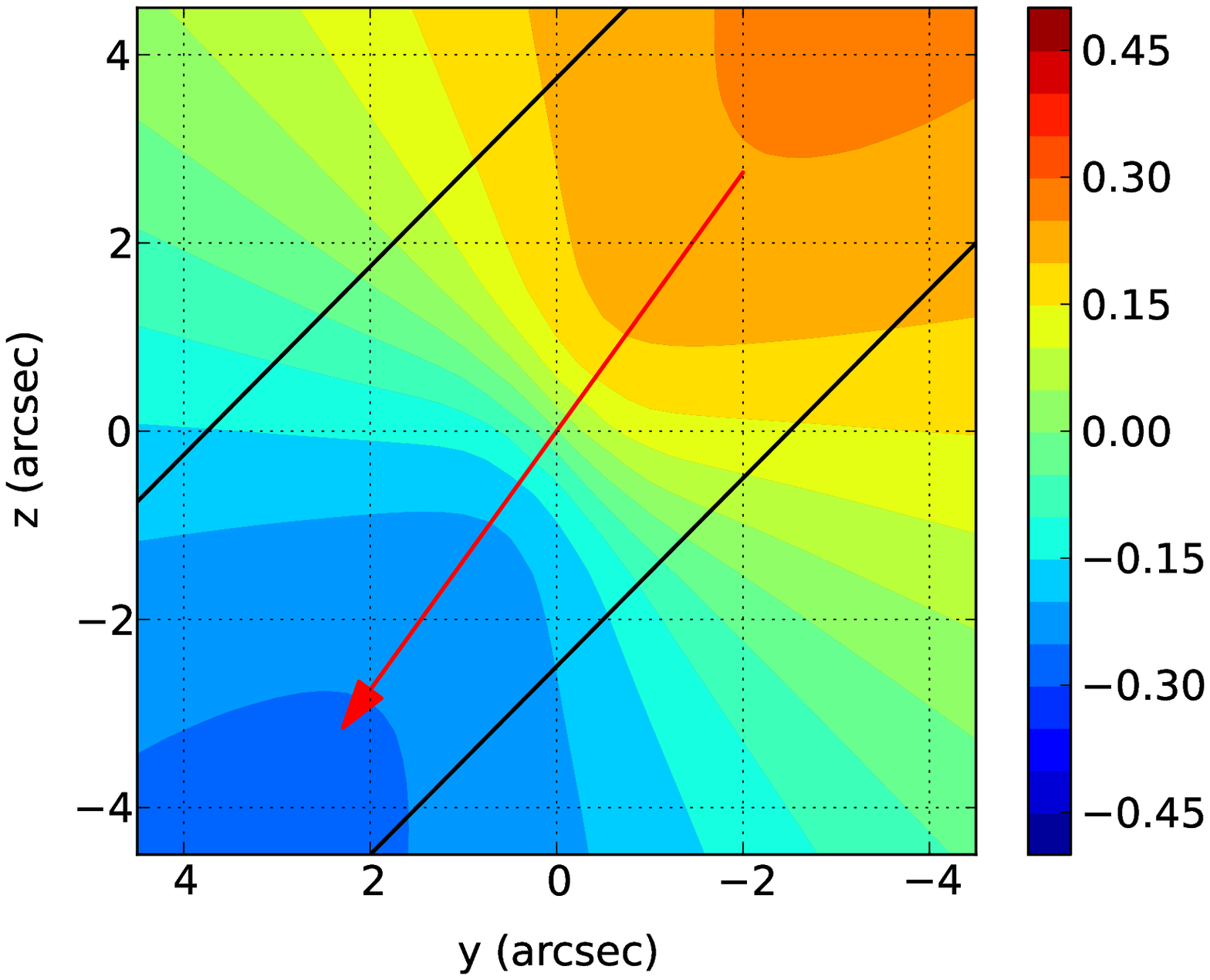}
\caption{Results of the model introduced in Sect.~\ref{compsec}
for the maps displayed in Fig.~\ref{fig12}. The red arrows indicate
the projection of the star axis on the sky plane.}
\label{fig14}
\end{figure*}

In order to assess quantitatively the validity of this interpretation,
we define a band running from south-east to north-west as shown in
Fig.~\ref{fig12}.  Calling
$\xi$ the sky coordinate running along the band, we display in
Fig.~\ref{fig11} (right) the dependence of $\Delta v$ on $\xi$ for
pixels contained inside the band. It reveals a complex structure with
successive bumps superimposed over a global increase, displaying
remarkable similarities between CO(1-0) and CO(2-1), implying that it
is likely to be real rather than an artifact of the analysis or
instrumental. As a model, we use the best fit introduced in
Sect.~\ref{compsec}, however with the temperature distribution and the
adjustment of the relative CO(1-0) to CO(2-1) normalization that were
introduced in Sect.~\ref{fluxsec}. As illustrated in Figs.~\ref{fig13}
and \ref{fig14}, the adopted model reproduces correctly the main
features of the  observed distributions: the width of the $\Delta v$
distribution and the  amplitude of the Doppler shift when moving from
south-east to north-west along  the $\xi$ axis. Again, as in the
preceding section, observations are well  reproduced by the model, not
only qualitatively but also quantitatively in  their general features.
This is only achieved by the hypothesis of a bipolar wind
structure and is inconsistent with the assumption of a spherical
outflow.

\section{Discussion}\label{descusec}

Applying a bipolar outflow model to analyze the spatially
resolved CO(1-0) and CO(2-1) spectra from EP Aqr results in an
excellent agreement of the model results with the observations. The
peculiar geometry of the star orientation, with the symmetry
axis nearly parallel to the line of sight, implies an
approximate circular symmetry when projected on the sky and gives the
illusion of a spherically symmetric gas distribution. However, when
correlated with the Doppler velocity distributions, it allows for a
transparent interpretation in terms of the spatial effective density,
each data-triple being associated with a single point in space
once the velocity of the radial wind is known. Under the hypotheses of
pure radial expansion and of rotation invariance about the star axis,
it is possible to reconstruct the space distribution of the effective
density.  As a first step, we have compared the observations with a
simple bipolar outflow model that had been developed earlier to
describe another AGB star, RS Cnc. A satisfactory description has been
obtained by adjusting the model parameters to best fit the CO(1-0) and
CO(2-1) data together. Several parameters defining the morphology and
the kinematics of the gas volume surrounding EP Aqr have been
evaluated this way with good confidence. Among these are the
orientation of the star axis, making an angle of ${\sim}13^\circ$ with
the line of sight and projected on the sky plane at ${\sim}144^{\circ}$
from north toward west. The velocity of the radial wind has been found
to increase from ${\sim}$2 km\,s$^{-1}$ at the star equator to 
${\sim}$10 km\,s$^{-1}$ at the poles. The flux of matter is also enhanced at the poles, 
however less than the radial wind, resulting in an effective density slightly enhanced
 in the equator region. This shows that an axi-symmetry 
such as that observed in EP Aqr (or \mbox{RS Cnc}, see figure 6 in Hoai et al. 2014) is more 
clearly revealed by velocity-resolved data. For instance infrared data obtained in 
the continuum, even at very high spatial resolution, could not reveal such 
an effect. Such a kind of morpho-kinematics 
could explain that asymmetries in AGB outflows are found 
preferentially through radio observations obtained at high spectral 
resolution (e.g. Castro-Carrizo et al. 2010). However, this characteristic 
of the model is partly arbitrary and its main justification is to give a good
description of the observations.  
We do not claim that the form adopted in the proposed model is
unique; on the contrary we are confident that other forms could have
been chosen with similar success. In particular, the hypothesis of
pure radial expansion retained in the model should not be taken as
evidence for the absence of rotation of the gas volume about the star
axis. Indeed, such rotation, if it were present, would be virtually 
undetectable as the resulting velocities, being nearly perpendicular
to the line of sight, would not produce any significant Doppler shift.

The assumptions made in formulating the model are always
approximations, some time very crude, of the reality. In particular,
evidence has been obtained for significant departures from the
symmetry with respect to the star equatorial plane assumed in the
model, revealing an excess of emission at northern star latitudes over
the whole longitudinal range at angular distances from the star
exceeding ${\sim}3''$. This feature, at least part of it, had already
been noted earlier (Winters et al. 2007) and thought to suggest the
existence of successive mass loss events. In other words a succession 
of mass loss events can be mimicked by a latitudinal variation of the 
flux of matter.

In general, while the proposed model has been very efficient at
describing the main features and general trends of the flux densities,
some significant deviations with respect to the observations have
also been found. The present data, in particular their spatial
resolution, prevent a more detailed statement about the nature of
these deviations. An important asset of the model is to illustrate the
efficiency of the method used to reconstruct the spatial morphology
and kinematics of the gas envelope. Once data of a
significantly better resolution will be available, e.g. from
ALMA, using the same methodology will reveal many of such details with
unprecedented reliability and precision.

The availability of measurements of the emission of two different
rotation lines of the same gas is extremely precious and has been
exploited as much as possible, based on the current data. The
flux ratio of two lines allows to evaluate the temperature
distribution of the gas in a region where the ratio between the
populations and emission probabilities of the two states vary rapidly,
which is the case for angular distances from the star probed by the
present observations. Most of the details of the model are irrelevant
to the value of the flux density ratio between the two lines, which
depends mostly on temperature under reasonable approximations (local
thermal equilibrium, absence of large turbulence, etc.). 
Through this property, evidence was found for a significant
enhancement of temperature at the star equator, therefore correlated
with the orientation of the bipolar flow and at variance with the
hypothesis of a pure radial dependence.  

A method allowing iterating
the wind velocity and the effective density in successive steps has
been sketched. When data with significantly improved spatial
resolution will be available, this will provide a very efficient
analysis tool. The peculiar orientation of EP Aqr makes such an
analysis much more transparent than in the general case where the star
axis is neither parallel nor perpendicular to the sky plane.

Furthermore, the good spectral resolution of the present observations
has made it possible to finely analyse the variations of the Doppler
velocity of the narrow line component across the sky map,
providing a sensitive test of the validity of the bipolar flow
hypothesis and a quantitative check of the inclination of the star
axis with respect to the line of sight. 

The present work leads us to prefer a bipolar, stationary, wind
model  over the spherical, variable, scenario that was proposed
earlier. AGB stars  have traditionally been assumed to be surrounded
by almost spherical  circumstellar shells. Deviations, sometimes
spectacular, from  spherical symmetry are observed in planetary
nebulae (PNs). Such deviations  are also observed  in post-AGB sources
and it is generally considered that  they arise during the phase of
intense mass loss, after the central stars  have evolved away from the
AGB. However, it has also been recognized  that the geometry  of
circumstellar shells around AGB stars may sometimes show an axi-symmetry
(e.g. X Her, Kahane \& Jura (1996); RS Cnc, Hoai et al. (2014)).  The
origin of this axi-symmetry has not been clearly identified:  magnetic
field, stellar rotation, or presence of a close binary companion may
play a role.  Recently, Kervella et al. (2015) reported the detection of a
companion to L2 Pup,  another M-type AGB star, with a disk seen almost
edge-on.  The case of EP Aqr is particularly interesting because, with
no detected Tc  and a $^{12}$C/$^{13}$C ratio $\sim$ 10, it seems to
be in a very early stage on the AGB. All these cases raise the
interesting  possibility that the deviations from sphericity observed
in PNs and in  post-AGB stars are rooted in their previous evolution
on the AGB.

\section{Conclusion}

We have revisited the EP Aqr data obtained by Winters et
al. (2007). We now interpret them in terms of an axi-symmetrical model
similar to that developed for RS Cnc by Hoai et al. (2014). The
expansion velocity varies smoothly with latitude, from
${\sim}2$\,km\,s$^{-1}$ in the equatorial plane to ${\sim}$10
km\,s$^{-1}$ along the polar axis, which for EP Aqr is almost
parallel to the line of sight. The mass loss rate is estimated to
${\sim}1.8$ \Mdot. The two stars look very similar, the
differences in the  observations being only an effect of the
orientations of their polar axis  with respect to the line of sight.
There is evidence for a temperature enhancement near the star equator
and a faster decrease along the polar axis than in the equatorial
plane. 

Both RS Cnc and EP Aqr show peculiar CO line
emission whose profiles consist of two components (Knapp et al. 1998,
Winters et al. 2003). This may suggest that the other sources with
such composite CO line profiles (e.g. X Her, SV Psc, etc.) might also
be interpreted with a similar model, rather than with successive winds
of different characteristics as has been proposed earlier.

\begin{acknowledgements} This work was done in the wake of an earlier
study, the results of which have been published (Winters et
al. 2007). We thank those who contributed to it initially, in the
phases of design, observations and data reduction for having
encouraged us to revisit the original analysis in new terms. Financial
support from the World Laboratory, from the French CNRS in the form of
the Vietnam/IN2P3 LIA and the ASA projects, from FVPPL, from PCMI,
from the Viet Nam National Satellite Centre, from the Vietnam National
Foundation for Science and Technology Development (NAFOSTED) under
grant number 103.08-2012.34 and from the Rencontres du Vietnam is
gratefully acknowledged. One of us (DTH) acknowledges support from the
French Embassy in Ha Noi. This research has made use of the SIMBAD and
ADS databases.

\end{acknowledgements}

\end{document}